\documentclass[10pt,a4paper]{article}
\usepackage[latin1]{inputenc}
\usepackage[T1]{fontenc}
\usepackage{amsmath}
\usepackage{amsfonts}
\usepackage{amssymb}
\usepackage{graphicx}
\begin{document}
\title{Renormalization of Very Special Relativity gauge theories}

\author{J. Alfaro \\
	Facultad de F\'\i sica, Pontificia Universidad Cat\'olica de Chile,\\
	Casilla 306, Santiago 22, Chile.\\
	jalfaro@uc.cl}

\maketitle

\begin{abstract}
	Mandelstam-Leibbrandt(ML) regularization of Very Special Relativity (VSR) amplitudes in momentum space depends on two fixed null vectors $n_\mu,\bar{n}_\mu$ besides external momenta. ML is known to preserve gauge invariance and naive power counting. The second null vector $\bar{n}_\mu$ destroys the $Sim(2)$ symmetry of the VSR model. We devise a systematic procedure to take the $\bar{n}_\mu->0$ limit to recover the lost $Sim(2)$ symmetry.
The procedure produces $Sim(2)$ and gauge invariant loop amplitudes. We show how to use this method to remove unphysical terms from the Non Relativistic Potential of Very Special Relativity Quantum Electrodynamics(VSR QED) with a massive photon. Then we compute the one loop renormalization of VSR QED. Finally we derive the anomalous magnetic moment of the electron, modified by neutrino and photon mass; and comments on the phenomenological implications.
\end{abstract}

\section{Introduction}
The Weinberg-Salam model(SM) is a very accurate unification of Weak and Electromagnetic interactions, that is being tested
at the LHC with a great precision. 

Remarkably, until today,
neither new particles nor new interactions have been discovered at the
LHC{\cite{w2}}. Some problems beyond the SM remain. Among the most important unsolved problems is neutrino's mass. SM postulates that
the neutrino is a massless particle, but the observed neutrino
oscillations{\cite{Langacker}} show that neutrinos are massive.

In a Lorentz invariant theory, we have to
introduce new particles and interactions in order to give masses to the
observed neutrinos through, for instance, the seesaw
mechanism{\cite{mohapatra}}.

Very Special
Relativity(VSR){\cite{CG1}} offers a new way to have massive neutrinos.  Instead of the 6 parameter Lorentz group, a 4
parameters subgroup($Sim(2)$) is assumed to be the symmetry of
Nature. $Sim(2)$ transformations change a fixed null four vector
$n_{\mu}$ at most by a scale factor, so ratios of scalar quantities containing
the same number of $n_{\mu}$ in the numerator as in the denominator are
$Sim(2)$ invariant, although  they are not Lorentz invariant. In
this context a new term in Dirac equation is allowed providing a VSR mass term for left handed
neutrinos{\cite{CG2}}.

Some time ago, we have introduced the SM with VSR{\cite{ja1}} (VSRSM), which accommodates the
same particles and interactions as  the SM, but neutrinos can have a VSR mass
without lepton number violation. As in the SM, the electron and the electron neutrino
form a $SU(2)_{L}$ doublet; therefore the VSR neutrino mass term will modify the QED of the electron.

Loop corrections in VSR Quantum Electrodynamics (VSR QED) have been computed in \cite{plb,AUniverse,AS1,AS2}. In these papers a infrared regulator of loop integrals introduced in \cite{plb} is employed, based on the calculation
of integrals using the Mandelstam-Leibbrandt(ML){\cite{Mandelstam}}
{\cite{Leibbrandt}}prescription introduced in {\cite{AML}}. We want to preserve gauge invariance, naive power counting and $Sim(2)$ symmetry. The first two properties are satisfied by
Mandelstam-Leibbrandt(ML) prescription. However ML introduces an additional
null vector $\bar{n}_{\mu}$ that breaks the $Sim(2)$ symmetry of the
model. In $\cite{plb}$ we explain how to trade $\bar{n}_{\mu}$ by
$n_{\mu}$ and some external vector(an external momentum in a loop integral). In this way we recover $Sim(2)$ invariance. However this
prescription fails to remove the infinities from the NRL(Please see Chapter 3 below).
Moreover, to preserve gauge invariance, we must modify the Feynman rules
$\cite{AUniverse}$.

In this  paper, we introduce a new and very simple way to recover $Sim(2)$ symmetry. We explain how to take the $\bar{n}_\mu->0$ limit in a systematic way. The procedure is very straightforward and powerful. Gauge symmetry and $Sim(2)$ symmetry are preserved all along, even in the presence of a massive photon.

Using this infrared regularization of $d$ dimensional integrals, we show how to recover the expected Yukawa potential in VSR QED.

The existence of unphysical potentials in the NRL of VSR theories with gauge invariant mass for the gauge field appears to be generic. It is also present in the gravitational potential in Very Special linear gravity(VSLG)\cite{ASan,ASanS}.The procedure presented in this paper remove the unphysical components of the gravitational potential.

Then we complete the one loop renormalization of VSR QED and get the anomalous magnetic moment of the electron, modified by the neutrino and photon mass.

The plan of the paper is the following: Chapter 1 contains the Introduction. Chapter 2 define the model, starting form VSRSM.
Chapter 3 introduces the lagrangian of Electrodynamics with a gauge invariant mass for the photon and study the Non Relativistic Limit(NRL) of this model. We show that physical inconsistencies appear. In Chapter 4 we introduce the new regularization. Chapter 5 study the renormalization of VSR QED. Chapter 6 contains the one loop renormalization of VSR QED. We compute photon and electron self energy and the corrections to the $e-e-\gamma$ vertex and show that the Ward identity for the photon self-energy and the Ward-Takahashi identity relating the $e-e-\gamma$ with the electron self energy is satisfied. We also calculate the vertex on shell and verify that is conserved. This is a very important test of the $Sim(2)$ limit proposed in this paper. In chapter 7 we compute the anomalous magnetic moment of the electron, corrected by a neutrino and photon mass. Chapter 8 study the implications of chapter 7 in the light of the most recent experimental data.
Chapter 9 draws some conclusions.

Appendix A contain the definitions of integrals used in the paper.Appendix B reports the vertex correction. It satisfies the Ward-Takahasi identity. Appendix C lists several useful identities among the integrals. Appendix D reports the small $\lambda$ behavior of various integrals, $\lambda=\frac{m_\gamma}{M_e}$, where $m_\gamma$ is the photon mass and $M_e$ is the electron mass. Appendix E presents a brief description of the method of traces to obtain the $\bar{n}_\mu->0$ limit.Appendix F contains the Feynman rules. 

\section{The model}

The leptonic sector of VSRSM consists of three $SU (2)$ doublets $L_{a} =
\left( \begin{array}{c}
	\nu^{0}_{aL}\\
	e^{0}_{aL}
\end{array} \right)$, where $\nu^{0}_{aL} = \frac{1}{2}  (1- \gamma_{5} )
\nu^{0}_{a}$ and $e^{0}_{aL} = \frac{1}{2}  (1- \gamma_{5} ) e^{0}_{a}$, and
three $SU (2)$ singlet $R_{a} =e^{0}_{aR} = \frac{1}{2}  (1+ \gamma_{5} )
e^{0}_{n}$. We assume that there is no right-handed neutrino. The index $a$
represent the different families and the index $0$ say that the fermionic
fields are the physical fields before breaking the symmetry of the vacuum.

In this letter we restrict ourselves to the electron family.

$m$ is the VSR mass of both electron and neutrino. 

After spontaneous symmetry breaking(SSB), the electron acquires a mass term
$M= \frac{G_{e} v}{\sqrt{2}}$, where $G_{e}$ is the electron Yukawa coupling
and $v$ is the VEV of the Higgs. Please see equation (52) of {\cite{ja1}}. The neutrino mass is not affected by SSB:$M_{\nu_{e}} =m$.

Restricting the VSRSM after SSB to the interactions between photon and
electron alone,  we get the VSR QED action.$\psi$ is the electron field.
$A_{\mu}$ is the photon field. We use the Feynman gauge.
\begin{eqnarray*}
	\mathcal{L}= \bar{\psi} \left( i \left( \not{D} + \frac{1}{2} \not{n} m^{2}
	(n \cdot D)^{-1} \right) -M \right) \psi - \frac{1}{4} F_{\mu \nu} F^{\mu
		\nu} - \frac{( \partial_{\mu} A_{\mu} )^{2}}{4} &  & \\
	D_{\mu} = \partial_{\mu} -i e A_{\mu} , & F_{\mu \nu} = \partial_{\mu}
	A_{\nu} - \partial_{\nu} A_{\mu} & 
\end{eqnarray*}
We see that the electron mass is $M_{e} = \sqrt{M^{2} +m^{2}}$, where $m$ is
the electron neutrino mass.

We present an extension of this model in the next section by adding a gauge invariant mass for the photon.

\section{VSR massive photon}

In {\cite{AS1}} we consider electrodynamics with a VSR mass for the
photon. The lagrangian for the photon with mass $m_{\gamma}$ is:
\begin{equation}
	\label{lagrangian} L_{gauge} = - \frac{1}{4} F_{\mu \nu} F^{\mu \nu} -
	\frac{1}{2} m_{\gamma}^2  (n^{\alpha} F_{\mu \alpha}) \frac{1}{(n \cdot
		\partial)^2}  (n_{\beta} F^{\mu \beta}) .
\end{equation}
Notice that this lagrangian is gauge invariant under the usual gauge
transformations:
\begin{equation}
	\delta A_{\mu} (x) = \partial_{\mu} \Lambda (x)
\end{equation}
\label{gi}

This is a very important property of a massive photon in VSR. It preserves
gauge invariance. Instead a Lorentz invariant mass for the photon breaks gauge
invariance.

In  Feynman gauge, we obtained the following photon propagator:
\begin{equation}
	\label{propph} \Delta_{\mu \nu} (p) = - \frac{i}{p^2 -
		m_{\gamma}^2}  \left[ g_{\mu \nu} + \frac{m_{\gamma}^2}{(n \cdot
		p)^2} n_{\mu} n_{\nu} - \frac{m_{\gamma}^2}{p^2 (n \cdot
		p)} (p_{\mu} n_{\nu} + p_{\nu} n_{\mu}) \right],
\end{equation} 

In the following sections we will use this propagator to compute the
Non-relativistic potential (NRL) of electrodynamics. 

\subsection{NRL of electrodynamics}

The Feynman rules where derived in {\cite{plb}}. We list them in Appendix F. At tree level we just need
the normal vertex:

\begin{figure}[h]
	\resizebox{239pt}{94pt}{\includegraphics{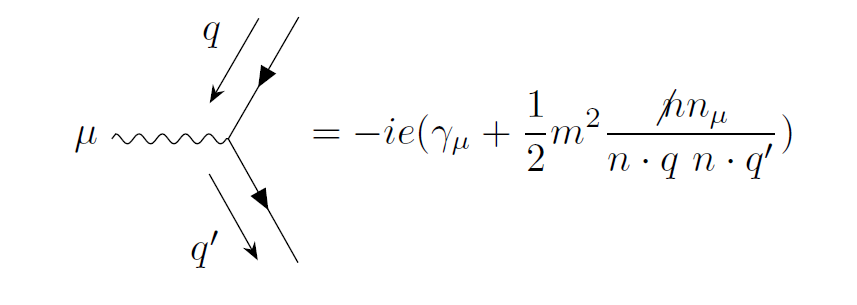}}
	\caption{$e - e - A_{\mu}$ vertex}
\end{figure}

We now consider $e - e$ scattering at tree level given by:
\begin{figure}[h]
	\resizebox{239pt}{94pt}{\includegraphics{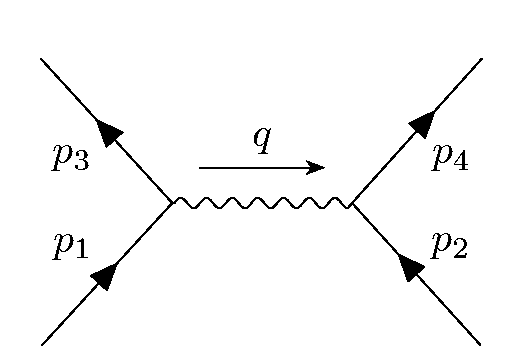}}
	\caption{$e+e->e+e\ \  {\rm scattering}$}
\end{figure}

That is:
\begin{eqnarray*}
	i\mathcal{M}= (- i e)^2 \bar{u}_3 \left( \gamma^{\mu} + \frac{1}{2} m^2
	\frac{\not{n} n^{\mu}}{n.p_1 n.p_3} \right) u_1 \Delta_{\mu \nu} (q)
	\bar{u}_4 \left( \gamma^{\nu} + \frac{1}{2} m^2 \frac{\not{n}
		n_{^{\nu}}}{n.p_2 n.p_4} \right) u_2, & q = p_1 - p_3 & 
\end{eqnarray*}
But the external legs are on-shell so
\begin{eqnarray*}
	q_{\mu} \bar{u}_3 \left( \gamma^{\mu} + \frac{1}{2} m^2 \frac{\not{n}
		n^{\mu}}{n.p_1 n.p_3} \right) u_1 = 0, & q_{\nu} \bar{u}_4 \left(
	\gamma^{\nu} + \frac{1}{2} m^2 \frac{\not{n} n_{^{\nu}}}{n.p_2 n.p_4}
	\right) u_2 = 0 & 
\end{eqnarray*}
Also:
\begin{eqnarray*}
	\bar{u}_3 \left( \gamma^{\mu} + \frac{1}{2} m^2 \frac{\not{n} n^{\mu}}{n.p_1
		n.p_3} \right) u_1 n_{\mu} n_{\nu} \bar{u}_4 \left( \gamma^{\nu} +
	\frac{1}{2} m^2 \frac{\not{n} n_{^{\nu}}}{n.p_2 n.p_4} \right) u_2 = &  & \\
	\bar{u}_3 \not{n} u_1 \bar{u}_4 \not{n} u_2 &  & 
\end{eqnarray*}
In the NRL, we have
\begin{eqnarray*}
	\bar{u}_3 \not{n} u_1 = 2 n_0 M \delta_{s_1 s_3} &  & \\
	\bar{u}_3 \gamma^{\mu} u_1 = 2 M \delta_{s_1 s_3} \delta^{\mu}_0 &  & 
\end{eqnarray*}
That is:
\begin{eqnarray*}
	i\mathcal{M}= i e^2 (2 M)^2 \frac{1}{q^2 - m_{\gamma}^2} \delta_{s_1 s_3}
	\delta_{s_2 s_4} \left( 1 + \frac{n_0^2 m_{\gamma}^2}{(n.q)^2} \right), &  &
	q = (0, \vec{q})
\end{eqnarray*}
Compare with the Born approximation 
\begin{eqnarray*}
	< p' | i T | p > & = - i \bar{V} (\vec{q}) 2 \pi \delta (E_{p'} - E_p), &
	\vec{q} = \vec{p}' - \vec{p}
\end{eqnarray*}
\begin{eqnarray*}
	\bar{V} (\vec{q}) = \frac{e^2}{\vec{q}^2 + m_{\gamma}^2} \left( 1 +
	\frac{m_{\gamma}^2}{(\hat{n} . \vec{q})^2} \right) &  & 
\end{eqnarray*}

\begin{eqnarray*}
	V (\vec{x}) = e^2 \int \frac{d^3 q}{(2 \pi)^3} \frac{e^{i \vec{q} .
			\vec{x}}}{\vec{q}^2 + m_{\gamma}^2} \left( 1 + \frac{m_{\gamma}^2}{(\hat{n}
		. \vec{q})^2} \right) &  & 
\end{eqnarray*}
The first term produced the expected Yukawa potential:
\begin{eqnarray*}
	V (\vec{x}) = \frac{e^2}{4 \pi} \frac{e^{- m_{\gamma} r}}{r} + e^2 \int
	\frac{d^3 q}{(2 \pi)^3} \frac{e^{i \vec{q} . \vec{x}}}{\vec{q}^2 +
		m_{\gamma}^2} \frac{m_{\gamma}^2}{(\hat{n} . \vec{q})^2} &  & 
\end{eqnarray*}
The second term is infrared divergent and need to be regularized. But for all regularizations the potential diverges for certain
values of $\frac{\vec{x} . \hat{n}}{r}$, for any $r$, which implies the
existence of infinite forces for certain angles, for any $r$, which is 	physically ruled out.

\section{A new regularization}

We can write:
\[ \int \frac{d^3 q}{(2 \pi)^3} \frac{e^{i \vec{q} . \vec{x}}}{\vec{q}^2 +
	m_{\gamma}^2} \frac{1}{(\hat{n} . \vec{q})^2} = - \int_{- \infty}^{\infty}
d x_0 \int \frac{d^4 q}{(2 \pi)^4} \frac{e^{- i q.x}}{q^2 - m_{\gamma}^2}
\frac{1}{(n.q)^2} \]
Let us compute
\[ U (x) = \int \frac{d^4 q}{(2 \pi)^4} \frac{e^{- i q.x}}{q^2 -
	m_{\gamma}^2} \frac{1}{(n.q)^2} \]
We want to preserve gauge invariance, naive power counting and $Sim(2)$ symmetry. The first two properties are satisfied by
Mandelstam-Leibbrandt(ML) prescription. However ML introduces an additional
null vector $\bar{n}_{\mu}$ that breaks the $Sim(2)$ symmetry of the
model. We have shown elsewhere $\cite{plb}$ how to trade $\bar{n}_{\mu}$ by
$n_{\mu}$ and some external vector; in this case $x_{\mu}$. However this
prescription fails to remove the infinities from the NRL.
Moreover, to preserve gauge invariance, we must modify the Feynman rules
$\cite{AUniverse}$.

Actually there is a much simpler way to recover $Sim(2)$ symmetry:To
take the limit $\bar{n}_{\mu} \rightarrow 0$. Then only one null vector will
remain: $n_{\mu}$.

Let us see how this limit can be implemented, using as an example,$U (x)$

M-L prescription is equivalent to the conditions $\cite{AML}$:
\begin{enumerate}
	\item $n.n = 0 = \bar{n} . \bar{n}$, $n. \bar{n} = 1$
	
	\item Scale invariance under $n_{\mu} \rightarrow \lambda n_{\mu},
	\bar{n}_{\mu} \rightarrow \lambda^{- 1} \bar{n}_{\mu}$.
	
	\item $f (x.x, n.x \bar{n} .x)$ must be regular at $n.x \bar{n} .x = 0$.
\end{enumerate}

Then $U(x)$ can be written as follows:
\begin{eqnarray*}
	U (x) = \int \frac{d^4 q}{(2 \pi)^4} \frac{e^{- i q.x}}{q^2 - m_{\gamma}^2}
	\frac{1}{(n.q)^2} = &  & \\
	(\bar{n} .x)^2 f (x.x, n.x \bar{n} .x) &  & 
\end{eqnarray*}
where the function $f (x.x, n.x \bar{n} .x)$ is uniquely determined by the three conditions we enumerated above. The
technique used to determine this function is explained in \cite{AML}.

To take the limit $\bar{n}_{\mu} \rightarrow 0$, we proceed in the following
way. Write $\bar{n}_{\mu} = \rho \bar{n}_{\mu}^{(0)}, n_{\mu} = \rho^{- 1}
n_{\mu}^{(0)}$, with

$\bar{n}_{\mu}^{(0)}, n_{\mu}^{(0)}$ satisfying condition 1.Then condition 1.
is satisfied for all $\rho$.

We define $\bar{n}_{\mu} \rightarrow 0$ by the limit $\rho \rightarrow 0$.

We get: $\lim_{\rho \rightarrow 0} \rho^2 (\bar{n}^{(0)} .x)^2 f (x.x, n^{(0)}
.x \bar{n}^{(0)} .x) = 0$.

That is, due to $Sim(2)$ symmetry, we must have:
\[ U (x) = \int \frac{d^4 q}{(2 \pi)^4} \frac{e^{- i q.x}}{q^2 - m_{\gamma}^2}
\frac{1}{(n.q)^2} = 0 \]
It follows that:
\[ \int \frac{d^3 q}{(2 \pi)^3} \frac{e^{i \vec{q} . \vec{x}}}{\vec{q}^2 +
	m_{\gamma}^2} \frac{1}{(\hat{n} . \vec{q})^2} = 0 \]
That is:
\begin{eqnarray*}
	V (\vec{x}) = \frac{e^2}{4 \pi} \frac{e^{- m_{\gamma} r}}{r} &  & 
\end{eqnarray*}
We get the expected Yukawa potential for a massive photon.

The same solution applies to the gravitational potential in Very Special linear gravity(VSLG)\cite{ASan,ASanS}.In VSLG the graviton propagator contains terms similar to the massive photon propagator equation (\ref{propph}). When we use it  to compute the classical gravitational potential between two masses, we get the same non-physical terms of the form:
\[ \int \frac{d^3 q}{(2 \pi)^3} \frac{e^{i \vec{q} . \vec{x}}}{\vec{q}^2 +
	m_{\gamma}^2} \frac{1}{(\hat{n} . \vec{q})^n}, n=2,4\]
The $Sim(2)$ limit of these integrals vanishes, so we recover a Yukawa-type gravitational potential.

\subsection{A more general integral}
Consider an arbitrary function $g$ and compute
\[ \int \frac{d^d q}{(2 \pi)^d} g (q^2, q.x) \frac{1}{(n.q)^a}\]
The M-L prescription using the method of $\cite{AML}$ implies
\[ \int \frac{d^d q}{(2 \pi)^d} g (q^2, q.x) \frac{1}{(n.q)^a} = (\bar{n}
.x)^a f (x.x, n.x \bar{n} .x) \]
for a unique $f (x.x, n.x \bar{n} .x)$, under the conditions:
\begin{enumerate}
	\item $n.n = 0 = \bar{n} . \bar{n}$, $n. \bar{n} = 1$
	
	\item Scale invariance under $n_{\mu} \rightarrow \lambda n_{\mu},
	\bar{n}_{\mu} \rightarrow \lambda^{- 1} \bar{n}_{\mu}$.
	
	\item $f (x.x, n.x \bar{n}.x)$ must be regular at $n.x \bar{n}.x = 0$.
\end{enumerate}
$x_{\mu}$ is an arbitrary vector.

To take the limit $\bar{n}_{\mu} \rightarrow 0$, write $\bar{n}_{\mu} = \rho
\bar{n}_{\mu}^{(0)}, n_{\mu} = \rho^{- 1} n_{\mu}^{(0)}$, with

$\bar{n}_{\mu}^{(0)}, n_{\mu}^{(0)}$ satisfying condition 1.Then condition 1.
is satisfied for all $\rho$.

We define $\bar{n}_{\mu} \rightarrow 0$ by the limit $\rho \rightarrow 0$.

We get: $\lim_{\rho \rightarrow 0} \rho^a (\bar{n}^{(0)} .x)^a f (x.x, n^{(0)}
.x \bar{n}^{(0)} .x) = 0$.

Thus:
\[ \int \frac{d^d q}{(2 \pi)^d} g (q^2, q.x) \frac{1}{(n.q)^a} = 0, a > 0 \]
It is clear that this result applies to loop integrals of the sort
{\cite{AML}}:
\begin{eqnarray}
	\int dp \frac{1}{[p^2 + 2 p.q - m^2 + i \varepsilon]^a}  \frac{1}{(n \cdot
		p)^b} = (- 1)^{a + b} i (\pi)^{\omega}  (- 2)^b \frac{\Gamma (a + b -
		\omega)}{\Gamma (a) \Gamma (b)}  (\bar{n} \cdot q)^b &  &  \nonumber\\
	\int_0^1 dtt^{b - 1}  \frac{1}{(m^2 + q^2 - 2 n \cdot q \bar{n} \cdot qt - i
		\varepsilon)^{a + b - \omega}}, & \omega = d / 2 &  \label{basic}
\end{eqnarray}
Therefore, the $Sim(2)$ limit is:
\[ \int dp \frac{1}{[p^2 + 2 p.q - m^2 + i \varepsilon]^a}  \frac{1}{(n \cdot
	p)^b} = 0, b > 0, q_{\mu} arbitrary \]

Taking derivatives in $q_{\mu}$, we get:
\[ \int dp \frac{1}{[p^2 + 2 p.q - m^2 + i \varepsilon]^a}  \frac{p_{\alpha_1}
	\ldots .p_{\alpha_n}}{(n \cdot p)^b} = 0, b > 0, q_{\mu} arbitrary
\]
Resuming, the $Sim(2)$ invariant regularization of any integral over
$p_{\mu}$, containing $\frac{1}{n.p}$ to any positive power must be evaluated
to zero.

It is clear that this procedure respects gauge invariance and $Sim(2)$ invariance.	

But how to proceed if $\gamma$ matrices are involved?

Let us consider an example:
\begin{eqnarray*}
	\int \frac{dp}{p^2 - m^2} \frac{\not{n}  \not{p}  \not{n}}{n.p} =
	\int \frac{dp}{p^2 - m^2} \frac{\left( 2 n.p - \not{p \not{n}}
		\right)  \not{n}}{n.p} = 2 \int \frac{dp}{p^2 - m^2} &  & 
\end{eqnarray*}
I can compute $p$ integral first, using ML:
\begin{eqnarray*}
	\int \frac{dp}{p^2 - m^2} \frac{\not{n}  \not{p}  \not{n}}{n.p} =
	\not{n} \gamma_{\mu}  \not{n} \int \frac{dp}{p^2 - m^2}
	\frac{p_{\mu}}{n.p} = \not{n} \gamma_{\mu}  \not{n} \int
	\frac{dp}{p^2 - m^2} \bar{n}_{\mu} &  & 
\end{eqnarray*}
the naive limit will give zero, but if we move $\not{n}$ to the right(or left) $\not{n}  \not{\bar{n}}  \not{n} = 2$ and
we get the same answer as before.

But, suppose that $\not{n}$ was already to the right. Consider:
\begin{eqnarray*}
	\int \frac{dp}{p^2 - m^2} \frac{ \not{p}  \not{n}}{n.p} = \int
	\frac{dp}{p^2 - m^2}  \not{\bar{n}}  \not{n} \rightarrow ? &  & 
\end{eqnarray*}
\emph{\bf Prescription:}
We move all $\not{n}$ to the right, pick up all $n.(p+Q)$ produced by this motion and use them to cancel as many $n.(p+Q)$ in the denominator as possible. Finally all remaining  $(n.(p+Q))^{-a},a>0$ are replaced by zero. Here $Q_\mu$ represents any vector different from $p_\mu$ (the integration variable) including the zero vector. Notice that $\frac{n.p}{n.(p+q)}=1$ because $\frac{n.p}{n.(p+q)}=1-\frac{n.q}{n.(p+q)}$ and the second term vanishes in the last step of the procedure.

According to this:
\begin{eqnarray*}
	\int \frac{dp}{p^2 - m^2} \frac{ \not{p}  \not{n}}{n.p} =0 
\end{eqnarray*}

and
\begin{eqnarray*}
	\int \frac{dp}{p^2 - m^2} \frac{\not{n}  \not{p}  \not{n}}{n.p} =
	2 \int \frac{dp}{p^2 - m^2} &  & 
\end{eqnarray*}

The rationale for this prescription is the following. We are interested in putting $\bar{n}_\mu=0$ to recover $Sim(2)$ invariance. But we cannot afford to loose gauge invariance. Gauge invariance appears in the form of Ward identities that the Feynman graphs must satisfy. If we write all graphs in a "canonical form" such as all $\not{n}$ to the right in all monomials(only one $\not{n}$ remains because $\not{n}.\not{n}=0$ ),the Ward identities that generally involves products with external momenta, will be satisfied for arbitrary values of $n_\mu$ and $\bar{n}_\mu$(To prove the Ward identity we do not need $n.n=\bar{n}.\bar{n}=0$,$n.\bar{n}=1$ when all $\not{n}$ are to the right of all $\not{\bar{n}}$ ). Then after evaluating $\bar{n}_\mu=0$, the Ward identity still will be satisfied in the surviving set of integrals defining the  graphs. This surviving set  define the $Sim(2)$ invariant gauge theory.

The prescription has a degree of arbitrariness. We could equally well use the convention of moving all $\not{n}$ to the left.

In the application to VSR QED we have checked whether this arbitrariness in the prescription produces ambiguities. We did not find any.

In Appendix E we present the method of traces to take the $\bar{n}_\mu=0$ limit. Using the trace method is obvious that the Ward identities are satisfied for the $\bar{n}_\mu=0$ sector. In VSR QED the trace method gives the same results as the one presented in this chapter.

In the next section we will apply this prescription to take the $\bar{n}_\mu->0$
limit($Sim(2)$ limit) to VSR QED. We will see that the answer is explicitly gauge invariant.

\begin{section}{Renormalization of VSR QED}	

In this chapter we follow \cite{Pokorski}. 

Since the $U ( 1 )$ gauge symmetry of the photon and electron remains intact, the whole
renormalized lagrangian of VSR QED is:
\begin{eqnarray*}
	\mathcal{L}_{R} =- \frac{1}{4} Z_{3} F_{\mu \nu} F^{\mu \nu} - \frac{1}{2}
	m_{\gamma}^{2} Z_{\gamma} (n^{\alpha} F_{\mu \alpha} ) \frac{1}{(n \cdot
		\partial )^{2}}  (n_{\beta} F^{\mu \beta} ) + &  & \\
	Z_{2} \bar{\psi} i \not{D}_{} \psi +Z_{\nu} \bar{\psi} \frac{i}{2} \not{n}
	m^{2}  (n \cdot D_{} )^{-1} \psi -Z_{_{0}} M \bar{\psi} \psi &  & \\
	D_{\mu} = \partial_{\mu} -i  e A_{\mu} &  & 
\end{eqnarray*}
	$Z$'s are renormalization constants.

$\mathcal{L}_{R}$ is invariant under renormalized gauge transformations:
\begin{eqnarray*}
	\psi' ( x ) =e^{i  \alpha ( x )} \psi ( x ) &  & \\
	A_{\mu}' ( x ) =A_{\mu} ( x ) + \frac{1}{e} \partial_{\mu} \alpha ( x )
\end{eqnarray*}
In perturbation theory, we write:
\begin{eqnarray*}
	\mathcal{L}_{R} =- \frac{1}{4} F_{\mu \nu} F^{\mu \nu} - \frac{1}{2}
	m_{\gamma}^{2}  (n^{\alpha} F_{\mu \alpha} ) \frac{1}{(n \cdot \partial
		)^{2}}  (n_{\beta} F^{\mu \beta} ) + &  & \\
	\bar{\psi} i \not{D} \psi + \bar{\psi} \frac{i}{2} \not{n} m^{2}  (n \cdot
	D_{R} )^{-1} \psi -M \bar{\psi} \psi + &  & \\
	- \frac{1}{4}  ( Z_{3} -1 ) F_{\mu \nu} F^{\mu \nu} - \frac{1}{2}
	m_{\gamma}^{2}  ( Z_{\gamma} -1 ) (n^{\alpha} F_{\mu \alpha} ) \frac{1}{(n
		\cdot \partial )^{2}}  (n_{\beta} F^{\mu \beta} ) + &  & \\
	( Z_{2} -1 ) \bar{\psi} i \not{D} \psi + ( Z_{\nu} -1 ) \bar{\psi}
	\frac{i}{2} \not{n} m^{2}  (n \cdot D_{R} )^{-1} \psi - ( Z_{_{0}} -1 ) M
	\bar{\psi} \psi &  & 
\end{eqnarray*}
and treat the counter terms involving $Z_3-1$,$Z_\gamma-1$,$Z_2-1$,$Z_\nu-1$ and $Z_0-1$ as perturbations.
\begin{subsection}{Renormalized photon mass}
	Let $\Pi_{\mu \nu}$ be the photon self energy.$\Pi_{\mu \nu}$ is symmetric, $Sim(2)$ invariant and satisfies the Ward identity,$p_\mu\Pi_{\mu \nu}=0$ . Therefore:
\begin{equation}
	\Pi_{\mu \nu} =A_{2} ( p_{\mu} p_{\nu} -p^{2} \eta_{\mu \nu} ) +A_{3} \left(
	\frac{p_{\mu} n_{\nu} +p_{\nu} n_{\mu}}{n.p} - \frac{p^{2} n_{\mu} n_{\nu}}{(
		n.p )^{2}} - \eta_{\mu \nu} \right)
\end{equation}
$Sim(2)$ invariance implies that $A_{i}$ is function of $p^{2}$ only.
Then, the inverse of the full propagators is:
\begin{eqnarray*}
	\Delta^{-1}_{\mu \nu} =- ( -p^{2} +m_{\gamma}^{2} ) \eta_{\mu \nu}
	-m_{\gamma}^{2} \frac{p^{2}}{( n.p )^{2}} n_{\mu} n_{\nu} +m_{\gamma}^{2}
	\frac{n_{\mu} p_{\nu} +n_{\nu} p_{\mu}}{n.p} - \Pi_{\mu \nu} = &  & \\
	\eta_{\mu \nu} ( p^{2} -m_{\gamma}^{2} +p^{2} A_{2} +A_{3} ) -A_{2}  p_{\mu}
	p_{\nu} + \frac{n_{\mu} p_{\nu} +n_{\nu} p_{\mu}}{n.p} ( m_{\gamma}^{2}
	-A_{3} ) + \frac{p^{2}}{( n.p )^{2}} n_{\mu} n_{\nu} ( A_{3} -m_{\gamma}^{2}
	) &  & 
\end{eqnarray*}
The full propagator is:
\begin{eqnarray*}
	\Delta_{\mu \nu} = \frac{1}{p^{2} -m_{\gamma}^{2} +p^{2} A_{2} +A_{3}}
	\left( \eta_{\mu \nu} + \frac{n_{\mu} p_{\nu} +n_{\nu} p_{\mu}}{n.p} \left(
	\frac{A_{3} -m_{\gamma}^{2}}{p^{2} ( A_{2} +1 )} \right) - \frac{p^{2}}{(
		n.p )^{2}} n_{\mu} n_{\nu} \left( \frac{A_{3} -m_{\gamma}^{2}}{p^{2} ( A_{2}
		+1 )} \right) \right) &  & \\
	+ \frac{A_{2}}{( A_{2} +1 ) ( p^{2} )^{2}} p_{\mu} p_{\nu} &  & 
\end{eqnarray*}
It is easy to check that the longitudinal part of the full propagator does not receive radiative corrections, which is required by the Ward identity.

Full propagator have a pole at $p^{2} =m_{R}^{2}$ ($m_{R}$ is the physical
photon mass) when
\begin{equation}
	m_{\gamma}^{2} =m_{R}^{2} ( 1+ ( A_{2} ( m_{R}^{2} ) + (Z_3-1) ) ) + (
	A_{3} ( m_{R}^{2} ) + ( Z_{\gamma}-1) m_{\gamma}^{2} )
\end{equation}
Near the pole:
\begin{eqnarray*}
	\Delta_{\mu \nu} \sim z_{\gamma}^{-1} \frac{1}{p^{2} -m_{R}^{2}} \left(
	\eta_{\mu \nu} - \frac{n_{\mu} p_{\nu} +n_{\nu} p_{\mu}}{n.p} +
	\frac{p^{2}}{( n.p )^{2}} n_{\mu} n_{\nu} \right) &  & \\
	z_{\gamma}^{-1} =1+ ( A_{2} ( m_{R}^{2} ) + ( Z_{3} -1 ) ) +m_{R}^{2} A_{2}'
	( m_{R}^{2} ) +A_{3}' ( m_{R}^{2} ) &  & 
\end{eqnarray*}
$z_{\gamma}$ is photon's wave function renormalization. 
\end{subsection}	
\begin{subsection}{Renormalized electron mass}
	Let $\Sigma ( p )$ the electron self energy.
Write:
\begin{eqnarray*}
	\Sigma ( p ) =A ( p^{2} ) \not{p} +B ( p^{2} ) \frac{\not{n}}{n.p} +C (
	p^{2} ) &  & 
\end{eqnarray*}	
The inverse of the full propagator is:
\begin{eqnarray*}
	S^{-1} = \not{p} -M- \frac{m^{2}}{2}  \frac{\not{n}}{n \cdot p} - \Sigma ( p
	) = &  & \\
	( 1-A ) \left( \not{p} - \frac{\bar{m}^{2}}{2} \frac{\not{n}}{n \cdot p} -
	\bar{M} \right) &  & \\
	\bar{m}^{2} = \frac{m^{2} +2B ( p^{2} )}{1-A ( p^{2} )} , \bar{M} =
	\frac{M+C ( p^{2} )}{1-A ( p^{2} )} &  & 
\end{eqnarray*}
Then:
\begin{eqnarray}
	S= \frac{1}{1-A ( p^{2} )} \frac{\left( \not{p} - \frac{\bar{m}^{2}}{2}
		\frac{\not{n}}{n \cdot p} + \bar{M} \right)}{p^{2} - \bar{m}^{2} ( p^{2} ) -
		\bar{M}^{2} ( p^{2} )} = &  &  \nonumber\\
	\frac{\left( \not{p} - \frac{\bar{m}^{2}}{2} \frac{\not{n}}{n \cdot p} +
		\bar{M} \right)}{p^{2} ( 1-A ( p^{2} ) ) -m^{2} -2B ( p^{2} ) - \frac{( M+C
			( p^{2} ) )^{2}}{1-A ( p^{2} )}} &  &  \label{poleelectron}
\end{eqnarray}
It has a pole when:
\begin{eqnarray}
	p^{2} = \bar{m}^{2} ( M_{R}^{2} ) + \bar{M}^{2} ( M_{R}^{2} ) =M_{R}^{2}  &  & 
\end{eqnarray}

Define

$K=p^{2} ( 1-A ( p^{2} ) ) -m^{2} -2B ( p^{2} ) - \frac{( M+C ( p^{2} )
	)^{2}}{1-A ( p^{2} )}$

The residue at the pole $p^2=M_R^2$ is
\begin{eqnarray*}
	z_e^{-1} =K' ( M_{R}^{2} ) =1-A ( M_{R}^{2} ) -M_{R}^{2} A' ( M_{R}^{2} ) -2B'
	( M_{R}^{2} ) - ( 2 \bar{M} ( M_{R}^{2} ) C' ( M_{R}^{2} ) +A' ( M_{R}^{2} )
	\bar{M} ( M_{R}^{2} )^{2} ) &  & 
\end{eqnarray*}
Introduce the notation $Q(M_R^2)=\bar{Q}$. Then in perturbation theory:
\begin{eqnarray*}
	z_e=1+ \bar{A} +M_{R}^{2} \bar{A}' +2  \bar{B}' +2M  \bar{C}' + \bar{A}' 
	M^{2} + \ldots &  & 
\end{eqnarray*}
$z_e$ is electron's wave function renormalization.
\end{subsection}
\end{section}
\section{One loop VSR QED with a gauge invariant photon mass}
In this section we apply our $Sim(2)$ regularization, based in the $\bar n_\mu=0$ limit to compute the renormalized  One particle Irreducible graphs at one loop in VSR QED with a gauge invariant photon mass. Most of the calculations have used FORM \cite{form}.

We will show that the Ward identity for the photon self energy is preserved and the Ward-Takahashi identity is satisfied. We will explicitly compute the counter terms in the On Shell Renormalization Scheme(OSR). All along the $Sim(2)$ symmetry is respected.

Finally the On Shell $\gamma-e-e$ vertex is evaluated. We verified that the renormalized vertex is conserved, in the OSR. To prove this is non-trivial. The gauge and $Sim(2)$ symmetry play a fundamental role.

Having done this, we were able to get a prediction for the anomalous magnetic moment of the electron, in the presence of a massive neutrino and a gauge invariant photon mass. It has log corrections in the photon mass which means that the model does not reduce to the one without a photon mass in the zero photon mass limit.These log corrections are actually interesting from the phenomenological point of view, because they enhance the very small contribution of the neutrino mass to the anomalous magnetic moment of the electron.
\begin{subsection}{Photon self energy}
	In this section we present the computation of the photon self-energy. 
	It is given by two graphs:
	
	\begin{figure}[h]
		\centering
		\includegraphics[width=0.2\textwidth]{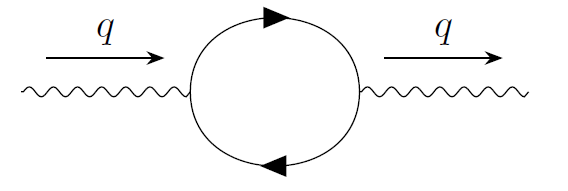}
		\includegraphics[width=0.2\textwidth]{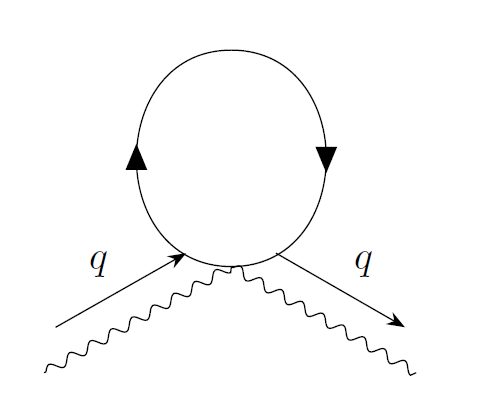}	
		\caption{Vacuum polarization one loop graphs}
		\label{Fig:vp}
	\end{figure}
\begin{eqnarray}
	i \Pi_{1 \mu \nu} =-(-ie)^{2} \int dp Tr (   \left[ \gamma_{\mu} +
	\frac{1}{2}  \frac{n_{\mu}   \not{n}  m^{2}}{n.(p+q)n.p} \right]  \frac{i
		\left( \not{p} +M- \frac{m^{2}}{2}  \frac{\not{n}}{n \cdot p} \right)}{p^{2}
		-M_{e}^{2} +i \varepsilon} &  & \nonumber\\
	\left[ \gamma_{\nu} + \frac{1}{2}  \frac{n_{\nu}   \not{n} 
		m^{2}}{n.(p+q)n.p}  \right] \frac{i \left( \not{p} + \not{q} +M-
		\frac{m^{2}}{2}  \frac{\not{n}}{n \cdot (p+q)} \right)}{(p+q)^{2} -M_{e}^{2}
		+i \varepsilon} )  &  & 
\end{eqnarray}
\begin{eqnarray}
	i \Pi_{2 \mu \nu} =e^{2 } i n_{\mu} n_{\nu}  \int dp \frac{1}{n.p n. (p+q)
		n. (p-q)} Tr (  \not{n} m^{2}  \frac{i \left( \not{p} +M-
		\frac{m^{2}}{2}  \frac{\not{n}}{n \cdot p} \right)}{p^{2} -M_{e}^{2} -m^{2}
		+i \varepsilon} )  &  & 
\end{eqnarray}

We use the new prescription to evaluate the diagrams. The result for the addition of the two graphs is:
\begin{eqnarray*}
	i  \Pi_{\mu \nu} =-4 e^2\int \frac{d^d p}{(2\pi)^d}   \frac{-2p_{\mu} p_{\nu} -p_{\mu} q_{\nu}
		-p_{\nu} q_{\mu} - \eta_{\mu \nu} ( M_{e}^{2} -p^{2} -p.q )}{( p^{2}
		-M_{e}^{2} +i \varepsilon ) ( ( p+q )^{2} -M_{e}^{2} +i \varepsilon )} &  & 
\end{eqnarray*}
Notice that some terms proportional to $m^{2}$ survive. They come from terms
created by the trace of the sort:$m^{2} n.p,m^{2} n. ( p+q )$. These terms
cancel the $\frac{1}{n.p},\frac{1}{n.(p+q)}$ so that after applying the $Sim( 2
)$ limit, they survive. These are just the terms we need to write the final
result entirely in terms of the physical electron mass, $M_{e}^{2} =M^{2} +m^{2}$, which is expected from unitarity.

This is the standard QED result, with the electron mass $M_{e}=\sqrt{M^2+m^2}$.

Write:
\begin{eqnarray*}
	\eta^{\mu \nu} \Pi_{\mu \nu} ( q ) = ( d-1 ) q^{2} \Pi ( q ) &  & 
\end{eqnarray*}
Then:
\begin{eqnarray*}
	( d-1 ) q^{2} \Pi ( q ) =-4i e^{2} \int \frac{d^d p}{(2\pi)^d}   \frac{-2p^{2} -2p.q-d (
		M_{e}^{2} -p^{2} -p.q )}{( p^{2} -M_{e}^{2} +i \epsilon ) ( ( p+q )^{2}
		-M_{e}^{2} +i \epsilon )} &  & 
\end{eqnarray*}
Define  $d=4- \epsilon , e \rightarrow e \mu^{\frac{\epsilon}{2}}$
\begin{eqnarray*}
	\Pi ( q^{2} ) =- \frac{\alpha}{\pi} \left[ \frac{1}{\varepsilon} \frac{2}{3}
	+ \left( \frac{1}{3} \log ( 4 \pi ) - \gamma \right) - \int_{0}^{1} d x 2x (
	1-x ) \log \left( \frac{M_{e}^{2} -x ( 1-x ) q^{2} -i  \varepsilon}{\mu^{2}}
	\right) \right] +o ( \varepsilon ) &  & 
\end{eqnarray*}
where $\alpha=\frac{e^2}{4\pi}$ is the fine structure constant.

In on-shell renormalization, we require that $m_{\gamma} =m_{R}$ and
$z_{\gamma} =1$. We get two conditions:
\begin{eqnarray*}
	- \Pi ( m_{\gamma}^{2} ) +Z_{3} -1+Z_{\gamma} -1=0 &  & \\
	- \Pi ( m_{\gamma}^{2} ) +Z_{3} -1-m_{\gamma}^{2} \Pi' ( m_{\gamma}^{2} ) =0
	&  & 
\end{eqnarray*}
\begin{eqnarray*}
	Z_{\gamma} -1=m_{\gamma}^{2} \Pi' ( m_{\gamma}^{2} ) =-2 \lambda^{2}
	\frac{\alpha}{\pi} \int_{0}^{1} d x  \frac{x^{2} ( 1-x )^{2}}{1-x ( 1-x )
		\lambda^{2} -i  \varepsilon} , & \lambda = \frac{m_{\gamma}}{M_{e}} & 
\end{eqnarray*}
a finite counter term.

\begin{eqnarray*}
	Z_{3} -1=- \frac{\alpha}{\pi} \left[ \frac{1}{\varepsilon} \frac{2}{3} +
	\left( \frac{1}{3} \log ( 4 \pi ) - \gamma \right) - \int_{0}^{1} d x 2x (
	1-x ) \log \left( \frac{M_{e}^{2} -x ( 1-x ) m_{\gamma}^{2} -i 
		\varepsilon}{\mu^{2}} \right) \right] &  & \\
	-2 \lambda^{2} \frac{\alpha}{\pi} \int_{0}^{1} d x  \frac{x^{2} ( 1-x
		)^{2}}{1-x ( 1-x ) \lambda^{2} -i  \varepsilon} &  & 
\end{eqnarray*}
\begin{equation}
\Pi_{OSR} ( q^{2} ) = \frac{\alpha}{\pi} \left[ \int_{0}^{1} d x 2x
( 1-x ) \log \left( \frac{M_{e}^{2} -x ( 1-x ) q^{2} -i 
	\varepsilon}{M_{e}^{2} -x ( 1-x ) m_{\gamma}^{2} -i  \varepsilon} \right)
\right] +2 \lambda^{2} \frac{\alpha}{\pi} \int_{0}^{1} d x  \frac{x^{2} (
	1-x )^{2}}{1-x ( 1-x ) \lambda^{2} -i  \varepsilon} 	
\end{equation}
\end{subsection}
\begin{subsection}{Electron Self Energy}
Here we calculate the electron self-energy. Again we have two graphs contributing to the 2-proper vertex.
See Figure(\ref{Fig:ese}).
\begin{figure}[h]
	\centering
	\includegraphics[width=0.2\textwidth]{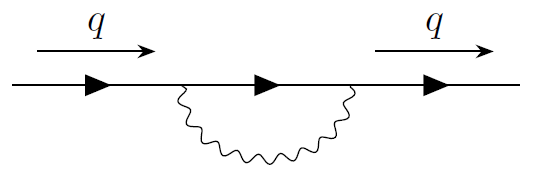}
	\includegraphics[width=0.2\textwidth]{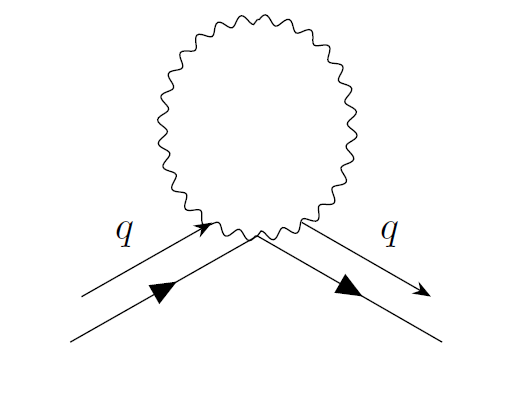}	
	\caption{Electron self energy one loop graphs. The second graph vanishes in Feynman gauge.}
	\label{Fig:ese}
\end{figure}

\begin{eqnarray}
	-i \Sigma_{1} (q) =(-ie)^{2} \int dp \left( \gamma_{\mu} + \frac{1}{2} m^{2}
	\not{n} n_{\mu}  \frac{1}{n.(p+q)}  \frac{1}{n.q} \right)  \frac{i \left(
		\not{p} + \not{q} +M- \frac{m^{2}}{2}  \frac{\not{n}}{n \cdot (p+q)}
		\right)}{(p+q)^{2} -M_{e}^{2} +i \epsilon} &  & \nonumber\\
	\left( \gamma_{\nu} + \frac{1}{2} m^{2} \not{n} n_{\nu}  \frac{1}{n.(p+q)} 
	\frac{1}{n.q} \right) \left( - \frac{n_{\mu}}{n.p}  \frac{n_{\nu}}{n.p} 
	\frac{im^{2}_{\gamma}}{p^{2} -m^{2}_{\gamma}} + \frac{n_{\mu} p_{\nu}
		+n_{\nu} p_{\mu}}{n.p(p^{2} -m^{2}_{\gamma} )} 
	\frac{im^{2}_{\gamma}}{p^{2}} - \frac{i \eta_{\mu \nu}}{p^{2}
		-m_{\gamma}^{2}} \right) &  & 
\end{eqnarray}
According to our Prescription, we move all $\not{n}$ to the right, pick up all $n.(p+Q)$ produced by this motion and use them to cancel as many $n.(p+Q)$ in the denominator as possible. Finally all remaining  $(n.(p+Q))^{-a},a>0$ are replaced by zero. Here $Q_\mu$ represents any vector different from $p_\mu$ (the integration variable) including the zero vector. Notice that $\frac{n.p}{n.(p+q)}=1$ because $\frac{n.p}{n.(p+q)}=1-\frac{n.q}{n.(p+q)}$.

Applying the $Sim(2)$ limit,we get:
\begin{eqnarray*}
	\Sigma_1 (p) = i e^2 
	2 m^2 \frac{\not{n}}{n.p} (S_3 - m_{\gamma}^2 I_3) + ((2 - d) (V_6 + S_3) +
	2 m_{\gamma}^2 I_3) \not{p} + M (d S_3 - 2 m_{\gamma}^2 I_3) &  & \\
	+ (Z_2 - 1) \not{p} + (Z_{\nu} - 1) \frac{1}{2} m^2 \frac{\not{n}}{n.p} -
	(Z_{_0} - 1) M &  & 
\end{eqnarray*}
For the definition of the functions used in this paper, please see Appendix A.

That is:
\begin{eqnarray*}
	A = (2 - d) (V_6 + S_3) + 2m_{\gamma}^2 I_3 + (Z_2 - 1) &  &
	\\
	B = 2 m^2 (S_3 - m_{\gamma}^2 I_3) + (Z_{\nu} - 1) \frac{1}{2} m^2 &  & \\
	C = M (d S_3 - 2 m_{\gamma}^2 I_3) - (Z_{_0} - 1) M &  & 
\end{eqnarray*}
On shell renormalization means:
$m$: Physical neutrino mass,$M_e^2 = M^2 + m^2$: Physical electron mass. That is:
\begin{eqnarray*}
	\bar{m}^2 (M_e^2) = m^2, & 2 B (M_e^2) + m^2 A (M_e^2) = 0 & \\
	\bar{M}_R (M_e^2) = M, & C (M_e^2) + A (M_e^2) M = 0 & 
\end{eqnarray*}
In addition, we must have:
\begin{eqnarray*}
	z_e = 1 &  & \\
	A(M_e^2)+ m^2 A'(M_e^2) + 2 B'(M_e^2) + 2 M C'(M_e^2) + 2 A'(M_e^2) M^2 = 0 & 
	& 
\end{eqnarray*}
These three conditions determines the three counter terms as the pole and finite part of the following expressions when $d->4$:
\begin{eqnarray*}
	Z_2 - 1 = (2-d) (V_6 (M_e^2) + S_3 (M_e^2)) - 2 m_{\gamma}^2 I_3 (M_e^2) - (2 B'
	(M_e^2) + m^2 A' (M_e^2) + 2 M C' (M_e^2) + 2 M^2 A' (M_e^2)) &  & \\
	Z_{\nu} - 1 = \frac{1}{m^2} (2 B' (M_e^2) + m^2 A' (M_e^2) + 2 M C' (M_e^2)
	+ 2 M^2 A' (M_e^2)) - 4 (S_3 (M_e^2) - m_{\gamma}^2 I_3 (M_e^2)) &  & \\
	Z_{_0} - 1 = (4 S_3 (M_e^2) - 2 m_{\gamma}^2 I_3 (M_e^2)  -
	\frac{1}{M} (2 B' (M_e^2) + m^2 A' (M_e^2) + 2 M C' (M_e^2) + 2 M^2 A'
	(M_e^2)) &  & 
\end{eqnarray*}
In the following computation of the on shell three vertex we assume on shell
renormalization of electron self energy.
\end{subsection}
\begin{subsection}{On shell vertex correction}
	In this subsection we discuss the 3 points proper vertex and verify the
	Ward-Takahashi identity. This is an important test of the gauge invariance of the infrared regulator. The one loop contribution to $\Gamma^{\mu} ( p' = p+q,p )$ consists of
	the addition of 3 graphs(Figure(\ref{Fig:3-vertex})):
	
	\begin{figure}[h]
		\centering
		\includegraphics[width=0.2\textwidth]{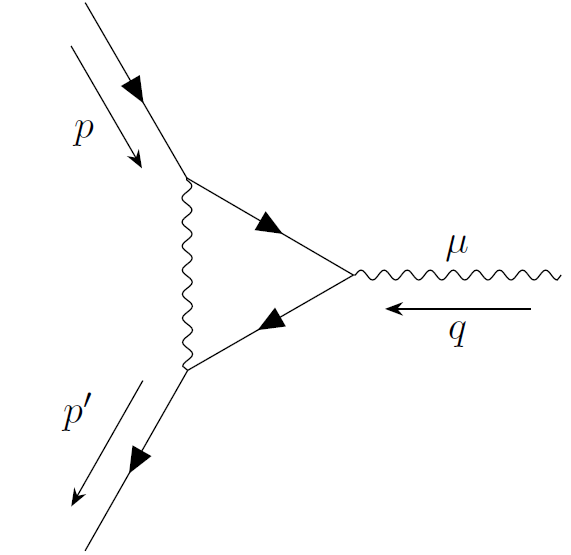}
		\includegraphics[width=0.2\textwidth]{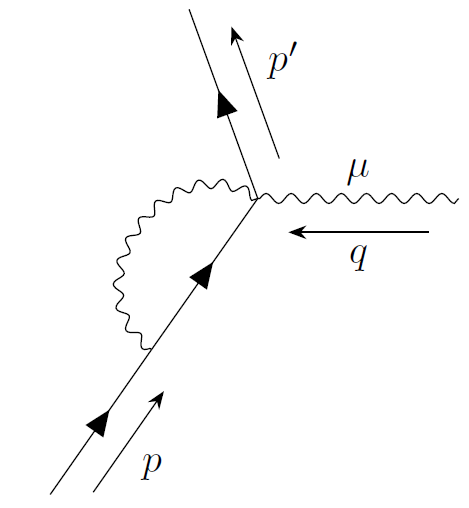}
		\includegraphics[width=0.2\textwidth]{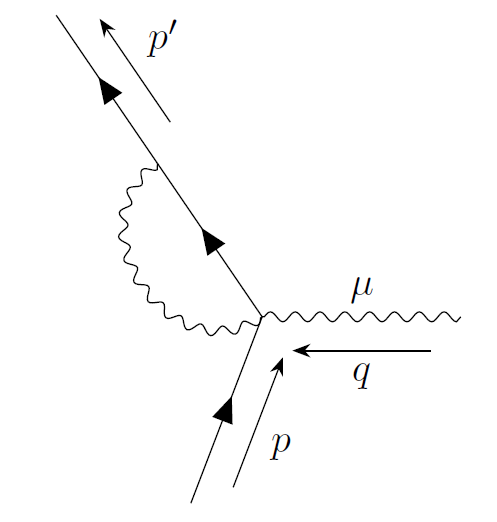}
		\caption{One loop contribution to the 3 points proper vertex}
		\label{Fig:3-vertex}
	\end{figure}

The contributions of the three graphs gives the vertex correction:
\begin{eqnarray}
	\delta \Gamma^{\mu}  (p+q,p) = \int dk \frac{i}{(k-p)^{2}} (-ie)^{2} \left[
	\gamma_{\nu} + \frac{1}{2} m^{2} \frac{n_{\nu} \not{n}}{n.(p+q)n.(k+q)} 
	\right]  \frac{\left( \not{k'} +M- \frac{m^{2}}{2}  \frac{\not{n}}{n \cdot
			k'} \right)}{k^{\prime 2} -M_{e}^{2} +i \varepsilon} &  & \nonumber\\
	\left[ \gamma_{\mu} + \frac{1}{2} m^{2} \frac{n_{\mu}   \not{n}}{n.(k+q)n.k}
	\right] \frac{\left( \not{k} +M- \frac{m^{2}}{2}  \frac{\not{n}}{n \cdot k}
		\right)}{k^{2} -M_{e}^{2} +i \varepsilon} \left[ \gamma_{\nu} + \frac{1}{2}
	m^{2} \frac{n_{\nu}   \not{n}}{n.p n.k}  \right] + &  & \nonumber\\
	(-ie)^{2} m^{2}  \int dk \frac{i}{(k-p-q)^{2}} \frac{1}{k^{2} -M_{e}^{2} +i
		\varepsilon} \frac{n_{\mu} \not{n}  n. ( k+p )}{n.p n. (p+q) n. (k-q)} + & 
	&\nonumber \\
	(-ie)^{2} m^{2}  \int dk \frac{i}{(k-p)^{2}} \frac{n_{\mu} \not{n}  n. (
		k+p+q )}{n.p n. (p+q) n. (k+q)} &  & \label{vertexcorrection}
\end{eqnarray}
Here $k'=k+q$. This vertex correction formally satisfy the Ward-Takahashi identity:
\begin{equation}
 - i q_{\mu} \Gamma^{\mu} (p + q, p) = S^{- 1} (p + q) - S^{- 1} (p) = - i
\left( \not{p} + \not{q} - M - \Sigma (p + q) - \not{p} + M + \Sigma (p)
\right) 
\end{equation}

Applying the $\bar n_\mu=0$ limit we obtain our result for the vertex correction. It is written in detail in Appendix B. It is well defined, because the integrals have been dimensionally regularized. We have verified that it satisfied the Ward Takahashi identity for any value of the parameters, including the non zero photon mass $m_\gamma$. This is a remarkable result of our prescription for the VSR integrals. For the first time we are able to incorporate a gauge invariant photon mass that preserve explicitly all the symmetries of the model.

Now we proceed to evaluate the vertex correction On shell.
That is:
\begin{eqnarray}
	\bar{u}(p+q))\delta\Gamma^\mu u(p)
\end{eqnarray}
with
\begin{eqnarray}
	\left( \not{p} - M - \frac{m^2}{2}  \frac{\not{n}}{n \cdot p} \right) u(p) = 0\\
	\bar{u}(p+q))	\left( \not{p}+\not{q} - M - \frac{m^2}{2}  \frac{\not{n}}{n \cdot p+n\cdot q} \right)=0
\end{eqnarray}
We have defined the Form Factors $G_2,G_3,F_3,F_1,F_2$.
\begin{eqnarray}
\bar{u} (p + q) \delta \Gamma^{\mu} (p + q, p) u (p)=R_{1\mu}+R_{2\mu}+\nonumber \\
\bar{u} (p + q) \left\{ G_2 \left[ - i \sigma_{\mu \nu} q_{\nu}\not{n}\right] + G_3 \not{n} Q_{\mu} + F_3 \not{n} \sigma_{\mu \nu} q_{\nu} \not{n} +
\tilde\gamma^\mu F_1 + F_2 i \frac{\sigma_{\mu \nu}}{2 M} q_{\nu} \right\}u(p) \label{ff}
\end{eqnarray}
where,
\begin{eqnarray}
	Q_\mu= q_\mu - q^2 \frac{n_\mu}{n.q}, & q.Q = 0,\\
\tilde{\gamma}^\mu=\gamma^{\mu} + \frac{m^2}{2} \frac{\not{n} n^{\mu}}{n.p (n.p + n.q)},\tilde\gamma^\mu q_\mu=0, on shell	
\end{eqnarray}
\begin{eqnarray}
	G_3 = \frac{im^2 n.q}{n.p (n.p + n.q)}  ((2 - d) \bar{T}_8 - (2 - d)
	\bar{T}_7 + \frac{\bar{T}_6 (1 - \frac{d}{2})}{2} + (2 - d) \bar{V}_4 +
	\frac{\bar{V}_3 (1 + \frac{d}{2}}{2})\\
	F_3 = \frac{m^2 M}{2 n.p (n.p + n.q)} \left( \left( 1 - \frac{d}{2} \right)
	\bar{T}_6 + \frac{d}{2} \bar{V}_3 \right)
\end{eqnarray}
\begin{eqnarray}
	G_2 = \frac{i m^2}{n.p} \left( - \frac{1}{2} \left( 1 - \frac{d}{2} \right)
	\bar{T}_6 - 2 \bar{V}_4 + \frac{1}{2} \left( 3 - \frac{d}{2} \right) \bar{V}_3
	\right) +\nonumber\\
	 \frac{i m^2}{n.p + n.q} \left( (d - 2) \bar{T}_7 + \frac{1}{2}
	\left( 1 - \frac{d}{2} \right) \bar{T}_6 - 2 \bar{V}_4 + \frac{1}{2} \left( 1
	+ \frac{d}{2} \right) \bar{V}_3 + 2 m_{\gamma}^2 \bar{V}_2 \right)
\end{eqnarray}
\begin{eqnarray}
	F_1 = - i (2 - d) \bar{V}_6 - i (8 - 4 d) \bar{T}_5 - i (d - 2) \bar{S}_4 + 2
	i m_{\gamma}^2 \bar{I}_4 - i (2 - d) (2 M^2 + m^2) \bar{T}_6\nonumber\\
	 - i (d - 2)
	\frac{n.q}{n. (p + q)} m^2 \bar{T}_7
	 - i (4 m^2 + 2 d M^2) \bar{V}_3 + 4 i m^2
	\frac{n.p}{n. (p + q)} m_{\gamma}^2 \bar{V}_2 + 2 i m^2 \frac{n.p}{n. (p + q)}
	m_{\gamma}^2 \bar{S}_1\nonumber\\ - i (4 - 2 d) q^2 \bar{T}_8 - i (2 + d) q^2 \bar{T}_7 +
	2 i q^2 \bar{T}_6 + Z_2 - 1\\
	F_2 = i M^2 \left( (6 d - 8 - d^2) \bar{T}_7 + \left( 8 - 5 d + \frac{d^2}{2}
	\right) \bar{T}_6 + (4 d - 16) \bar{V}_4 + \left( 5 d - 8 - \frac{d^2}{2}
	\right) \bar{V}_3 \right)
\end{eqnarray}

The two terms $R_{1\mu}$ and $R_{2\mu}$ are in general, non-zero. We find:
\begin{equation}
	R_{1 \mu} = q_{\mu} \bar{u} (p + q) u (p) \left( 2 - \frac{3 d}{2} +
	\frac{d^2}{4} \right) i M (\bar{V}_5 - \bar{V}_6 + \bar{S}_4 - \bar{S}_3)
\end{equation}
which is zero on shell since $\bar{V}_5 = \bar{V}_6$ and $\bar{S}_3 =\bar{S}_4$.Please remember that for integral $Q(p,q)$ we use the notation $\bar Q=Q(p^2=M_e^2,q)$.

Actually $R_{1\mu}$ appears also in standard QED, so it is not a surprise that it cancels on shell.

$R_{2\mu}$ is trickier. It is absent in standard QED. It is divergent in $d=4$.
\begin{equation}
R_{2\mu}=i\bar{u} (p + q)\frac{\not{n} n_\mu}{n.p (n.p+n.q)} u(p)m^2(\left( \frac{d}{2} - 1 \right) \bar{V}_6 + \left( \frac{d}{2} - 3 \right)
\bar{S}_3 + m_{\gamma}^2 \bar{I}_3)
\end{equation}
To cancel this term from the renormalized three vertex, we must add the vertex counter term.
The VSR QED three vertex has two different counter terms, in contrast to QED where the three vertex has one counter term.

From the lagrangian we can read the counter term for the 3-point function.

. We have to identify terms linear in $A_{\mu}$ in
\begin{eqnarray*}
	(Z_2 - 1) \bar{\psi} i \not{D} \psi + (Z_{\nu} - 1) \bar{\psi} \frac{i}{2}
	\not{n} m^2 (n \cdot D)^{- 1} \psi &  & 
\end{eqnarray*}
That is:
\begin{eqnarray*}
	(Z_2 - 1) \gamma^{\mu} - (Z_{\nu} - 1) \frac{1}{2} m^2 \frac{n_{\mu}
		\not{n}}{n.p n. (p + q)} = &  & \\
	(Z_2 - 1) \left( \gamma^{\mu} + \frac{1}{2} m^2 \frac{n_{\mu} \not{n}}{n.p
		n. (p + q)} \right) - \frac{1}{2} m^2 \frac{n_{\mu} \not{n}}{n.p n. (p + q)}
	((Z_{\nu} - 1) + (Z_2 - 1)) &  & 
\end{eqnarray*}
The first counter term renormalizes the coupling of the photon to the electric current, i.e. $F_1$.
The counter term that affects $R_{2\mu}$ is the last one. Due to charge conservation the addition of the counterterm $- \frac{1}{2} m^2 \frac{n_{\mu} \not{n}}{n.p n. (p + q)} ((Z_{\nu} - 1) +
(Z_2 - 1))$ must cancel $R_{2\mu}$. It is easy to check that it does.

It remains to show that on shell renormalization implies that $e$ is the physical electron charge. To do this we must show that $F_1(q^2=0)=1$.

Let us see how it goes in VSR QED.
\subsection{$F_1 (0)$}

Near on shell, the Ward-Takahashi identity is:
\begin{eqnarray*}
	- i k_{\mu} \Gamma^{\mu} (p + k, p) = S^{- 1} (p + k) - S^{- 1} (p) = 
	z_e^{- 1} \left[ \not{k} + \frac{m^2}{2} \frac{\not{n} n.k}{(n.p)^2} \right] &
	& 
\end{eqnarray*}
But the renormalization of the electric charge is:
\begin{equation}
	\Gamma^{\mu} (p, p) = i F_1 (0) \left( \gamma^{\mu} + \frac{m^2}{2}
\frac{\not{n} n^{\mu}}{(n.p)^2} \right)
\end{equation}
Into the Ward-Takahashi identity:
\begin{eqnarray*}
	F_1 (0) \left( \not{k} + \frac{m^2}{2} \frac{\not{n} n.k}{(n.p)^2} \right) =
	z_e^{- 1} \left[ \not{k} + \frac{m^2}{2} \frac{\not{n} n.k}{(n.p)^2} \right] &
	& \\
	F_1 (0) = z_e^{-1}
\end{eqnarray*}
Using various identities that we list in Appendix C, we are able to show that this equality holds in our infrared ($\bar{n}^\mu=0$ limit)and ultraviolet (Dimensional Regularization) regularization.
But, on shell renormalization of the electron propagator imposes $z_e=1$. Therefore $F_1(0)=1$ and $e$ represents the physical electron charge.

We are prepared now to extract the anomalous magnetic moment of the electron, in next section.
\end{subsection}
\section{Anomalous magnetic moment of the electron}
 In the Non-Relativistic(NR) limit we get Table 1, keeping terms that are at most linear in $q_{\mu}$.
 \begin{table}[h]
 	\begin{center}
 		\begin{tabular}{lp{.8\linewidth}}
 			NR limit & Form factor\\[5pt]
 			$2M_{e} \varphi^{\uparrow}_{s} \varphi_{s} A_{0}$ & $F_{1}$(0)\\
 			$\frac{3m^{2}}{4M^{2}} i  \varepsilon_{i j k} \varphi^{\uparrow}_{s}
 			\sigma^{i} \varphi_{s'} \hat{n}_{j} q_{k} A_{0}$ & $F_{1}$(0)\\
 			$i  \varepsilon_{i j k} q_{j} \varphi^{\uparrow}_{s} \sigma^{k} \varphi_{s'}
 			A_{i}$ & $F_{1}$(0)\\
 			$-2 i n_{0}  M \varepsilon_{i j k} \varphi^{\uparrow}_{s} \sigma^{k}
 			\varphi_{s'} q_{j} A_{i}$ & $G_{2} ( 0 )$\\
 			$-i  \varepsilon_{i j k} n_{k} \frac{m^{2}}{M} \varphi^{\uparrow}_{s} \hat{n}
 			. \vec{\sigma} \varphi_{s'} q_{j} A_{i}$ & $G_{2} ( 0 )$\\
 			$i ( 2M \varepsilon_{i j k}  n_{k} \varphi^{\uparrow}_{s} \sigma^{j}
 			\varphi_{s'} +2M_{e}  i n_{i} \varphi^{\uparrow}_{s} \varphi_{s'} ) A_{0} 
 			q_{i}$ & $G_{2} ( 0 )$\\
 			$2M_{e}  n_{0} \varphi^{\uparrow}_{s} \varphi_{s'} Q_{\mu} A^{\mu}$ & $G_{3} ( 0 )$\\
 			$( -4M_{e} \varepsilon_{i j k} n_{k} \varphi^{\uparrow}_{s} \vec{n} .
 			\vec{\sigma} \varphi_{s'} +4M_{e} n_{0}^{2} \varepsilon_{i j k}
 			\varphi^{\uparrow}_{s} \sigma^{k} \varphi_{s'} ) q_{j} A_{i}$ & $F_{3} ( 0 )$\\
 			$4M_{e}  n_{0} \varepsilon_{i j k} n_{j} \varphi^{\uparrow}_{s} \sigma^{k}
 			\varphi_{s'} A_{0} q_{i}$ & $F_{3} ( 0 )$\\
 			$i  \varepsilon_{i j k} \varphi^{\uparrow}_{s} \sigma^{k} \varphi_{s'} A_{i}
 			q_{j}$ & $F_{2} ( 0 )$\\
 			$-i \frac{m^{2}}{2M^{2}} \varepsilon_{i j k} \hat{n}_{j}
 			\varphi^{\uparrow}_{s} \sigma^{k} \varphi_{s} A_{0} q_{j}$ & $F_{2} ( 0 )$
 		\end{tabular}
 	\end{center}  
 	\caption{In the right column we list  the form factor. In the left column
 		we have the NR limit of the matrix element accompanying the form factor in
 		(\ref{ff}).All form factors are evaluated at $q_{\mu} =0$. Here $A_{0}$ is the electric
 		potential and $A_{i}$ is the vector potential.$\varphi_{s'}$ is  a two
 		dimensional constant vector that corresponds to the NR limit of the Dirac
 		spinors.
 	}
 \end{table}

 Then the anomalous magnetic moment is:
\begin{equation}
 a_e =F_{2} ( 0 ) -2n_{0} M G_{2} ( 0 ) -4i M_{e} n_{0}^{2} F_{3} ( 0) 
\end{equation}
In Appendix D we list the integrals we used to calculate the form factors.$\sim$ means the small $\lambda$ limit.
We get:
\begin{eqnarray}
F_{2} ( 0 ) =i 4M^{2} ( \bar{V}_{3} ( 0 ) - \bar{T}_{6} ( 0 ) )\sim \frac{\alpha}{2 \pi} \frac{M^{2}}{M_{e}^{2}}\\
G_2(0)=i m^{2} \frac{1}{n.p} ( 4 \bar{S}_{2} ( 0 ) -2 \bar{V}_{3} ( 0 ) +2
\bar{T}_{7} ( 0 ) +2m_{\gamma}^{2} \bar{V}_{2} ( 0 ) )\sim - \frac{m^{2}}{M_{e}^{2}} \frac{\alpha}{2 \pi n_{0} M_{e}} \left( \log (
\lambda ) - \frac{3}{4} \right)\\
F_3(0)=m^{2} M \frac{1}{( n.p )^{2}} \left( - \frac{1}{2} \bar{T}_{6} ( 0 ) +
\bar{V}_{3} ( 0 ) \right)=i \frac{m^{2}}{M_{e}^{2}} \frac{M}{n_{0}^{2} M_{e}^{2}} \frac{\alpha}{4 \pi}
\left( \frac{1}{2} \log ( \lambda ) + \frac{1}{4} \right)
\end{eqnarray}

That is:
\begin{equation}
a_e= \frac{\alpha}{2 \pi} \left( 1+
	\frac{m^{2}}{M_{e}^{2}} \left( 3 \log ( \lambda ) -2 \right) \right)
\end{equation}

\section{Phenomenology}

From Particle Data Group\cite{pdg}:
\begin{equation}
m_{\gamma}  < 3 \times 10^{-27} eV / {c}^{2} 
\end{equation}

The current experimental value and uncertainty for $a_e$ is\cite{ammexp}
$a_{\text{e}} =0.00115965218073 ( 28 )$

The $\alpha^{5}$ QED prediction is\cite{ammtheo}:$a_e=0.00115965218164(764)$.

Assuming that the difference between QED prediction and experimental value is due to the mass of the photon, we get:
\begin{equation}
\frac{m^{2}}{M_{e}^{2}} \left( -3 \log ( \lambda ) +2 \right) \sim 7.9\times 10^{-10}
\end{equation}
Using the current bound on the photon mass, we get
\begin{equation}
\frac{m}{M_{e}} \leq 1.9 \times 10^{-6}
\end{equation}

This value puts the electron neutrino mass around 1eV or less. 
Remarkable the most recent electron anti neutrino  mass bound is $m_\nu<0.8 eV c^-2$\cite{katrin}.

But $m_{\gamma}$
could be smaller, implying a smaller electron neutrino mass.

In fact, the most stringent bound on neutrino masses come from Cosmology\cite{cosmo1,cosmo2}
\begin{equation}
\sum m_i<0.12 eV
\end{equation}
If $m_\nu \sim 0.12 eV$, we get $\lambda\sim e^{-4500}$, which is a tiny but non-zero photon mass.

Recently the Fermilab Muon $g-2$  experiment \cite{fermilab} confirmed the measurement at Brookhaven National Laboratory\cite{brook} to produce a world average for the anomalous magnetic moment of the muon:
\begin{equation}
	a^{experimental}_\mu=116 592 061(41) \times  10^{-11}
\end{equation}
which differ from the SM theory prediction \cite{SMmuon}
\begin{equation}
	a^{SM}_\mu=116 591 810(43)  \times  10^{-11}
\end{equation}
by $4.2\sigma$.This tension could be a signal of Physics beyond the SM.
It may well be that corrections from massive neutrinos and massive photon
to $a_\mu$ may play a role in relaxing the tension. The $\bar{n}_\mu->0$ limit explained in this paper
can be used to explore this possibility.

\section{Conclusions}

In this paper we have introduced a prescription to obtain the 
$Sim(2)$ limit of VSR graphs, infrared regularized using the ML regularization. In ML, besides the $n_\mu$ null vector of VSR theories, a second null vector $\bar{n}_\mu$ is required. 
ML preserves naive power counting and gauge invariance, but destroys the $Sim(2)$ symmetry.

To recover the $Sim(2)$ symmetry we take the limit $\bar{n}_\mu->0$. For scalar integrals this limit is well defined and straightforward.

In the presence of $\gamma$ matrices we need to introduce a convention. In this paper we used the convention to move all
$\not{n}$ to the right and then take $\bar{n}_\mu->0$. In Appendix E we present the method of traces to compute $\bar{n}_\mu->0$. In VSR QED both methods give the same answer.

 $\bar{n}_\mu->0$ selects a subset of integrals defining the $Sim(2)$ regularized  Feynman graphs. The Ward identities are satisfied among graphs belonging to this subset.
 
 Using this prescription we proceed to compute the one loop renormalization of VSR QED with a gauge invariant photon mass. We have shown that the photon self energy is transverse and that the Ward- Takahashi identity is satisfied for the $Sim(2)$ invariant  graphs.
 
 Then we computed the three vertex on shell and checked that is conserved; we  extracted the form factors and compute the anomalous magnetic moment of the electron, corrected by non-zero neutrino and photon mass. 
 
 Using the most recent data for the photon mass , we obtain a bound for the electron neutrino mass.
 
 The $Sim(2)$ regulator we defined in this paper is very simple and reduce all integrals to standard dimensionally regularized integrals. It is universal, i.e, it apply to all VSR graphs, because the zero vector is an invariant vector for $Sim(2)$ too. It can be applied to all sorts of VSR theories, not just VSR QED. 
 We are guarantied to preserve the gauge invariance of the model.

\begin{section}{Acknowledgments}

J.A.acknowledges the partial support of the Institute of
Physics PUC and Fondo Gemini Astro20-0038.
\end{section}

\section{Appendix A}
\begin{eqnarray}
S_{1} = \int d k \frac{1}{( k-p )^{2}} \frac{1}{( k-p )^{2} -m_{\gamma}^{2}}
\frac{1}{k^{2} -M_{e}^{2}} \frac{1}{( k+q )^{2} -M_{e}^{2}} \\
\int d k \frac{k_{\mu}}{( k-p )^{2}} \frac{1}{( k-p )^{2} -m_{\gamma}^{2}}\frac{1}{k^{2} -M_{e}^{2}} \frac{1}{( k+q )^{2} -M_{e}^{2}}=V_1 p_\mu+V_2 q_\mu \\
\int d k \frac{k_{\mu} k_{\nu}}{( k-p )^{2}} \frac{1}{( k-p
	)^{2} -m_{\gamma}^{2}} \frac{1}{k^{2} -M_{e}^{2}} \frac{1}{( k+q )^{2}
	-M_{e}^{2}}=T_{1} \eta_{\mu \nu} +T_{2} p_{\mu} p_{\nu} +T_{3} ( p_{\mu} q_{\nu} +p_{\nu}
q_{\mu} ) +T_{4} q_{\mu} q_{\nu}\\
S_{2} = \int d k \frac{1}{( k-p )^{2} -m_{\gamma}^{2}} \frac{1}{k^{2}
	-M_{e}^{2}} \frac{1}{( k+q )^{2} -M_{e}^{2}}\\
\int d k \frac{k_{\mu}}{( k-p )^{2} -m_{\gamma}^{2}} \frac{1}{k^{2}
	-M_{e}^{2}} \frac{1}{( k+q )^{2} -M_{e}^{2}}=V_{3} p_{\mu} +V_{4} q_{\mu}\\
\int d k \frac{k_{\mu} k_{\nu}}{( k-p )^{2} -m_{\gamma}^{2}} \frac{1}{k^{2}
	-M_{e}^{2}} \frac{1}{( k+q )^{2} -M_{e}^{2}}=T_{5} \eta_{\mu \nu} +T_{6} p_{\mu} p_{\nu} +T_{7} ( p_{\mu} q_{\nu} +p_{\nu}
q_{\mu} ) +T_{8} q_{\mu} q_{\nu}\\
S_{3} = \int d k \frac{1}{( k-p )^{2} -m_{\gamma}^{2}} \frac{1}{k^{2}
	-M_{e}^{2}}\\
S_{4} = \int d k \frac{1}{( k-p' )^{2} -m_{\gamma}^{2}} \frac{1}{k^{2}
	-M_{e}^{2}}\\
\int d k \frac{k_{\mu}}{k^{2} -m_{\gamma}^{2}} \frac{1}{( k+p' )^{2}
	-M_{e}^{2}} =V_{5} p'_{\mu}\\
\int d k \frac{k_{\mu}}{k^{2} -m_{\gamma}^{2}} \frac{1}{( k+p )^{2}
	-M_{e}^{2}} =V_{6} p_{\mu}\\
I_{3}(p^2) = \int d k \frac{\frac{1}{( k-p )^{2}}}{( ( k-p )^{2}
	-m_{\gamma}^{2} ) ( k^{2} -M_{e}^{2} )}\\
I_{4}(p'^2) = \int d k \frac{\frac{1}{( k-p )^{2}}}{( ( k-p )^{2} -m_{\gamma}^{2} )( ( k+q )^{2} -M_{e}^{2} )}\\
\end{eqnarray}

\section{Appendix B}
In this Appendix we write the result for the vertex correction
defined in equation  (\ref{vertexcorrection}) in Form notation. Notice that mg=$m_\gamma$.
\begin{verbatim}
  deltaGamma(mu) =
+ g_(1,n) * (  - n(mu)*i_*m^2*[np]^-1*[np+nq]^-1*S4 - n(mu)*i_*m^2*
[np]^-1*[np+nq]^-1*S3 + n(mu)*i_*m^2*[np]^-1*[np+nq]^-1*mg^2*I4 + 
n(mu)*i_*m^2*[np]^-1*[np+nq]^-1*mg^2*I3 - 2*p(mu)*i_*m^2*[np]^-1*V3
+ 2*p(mu)*i_*m^2*[np]^-1*mg^2*V1 - 2*p(mu)*i_*m^2*[np+nq]^-1*V3 + 2*
p(mu)*i_*m^2*[np+nq]^-1*mg^2*V1 - 2*q(mu)*i_*m^2*[np]^-1*S2 - 2*q(mu)
*i_*m^2*[np]^-1*V4 + 2*q(mu)*i_*m^2*[np]^-1*mg^2*S1 + 2*q(mu)*i_*m^2*
[np]^-1*mg^2*V2 - 2*q(mu)*i_*m^2*[np+nq]^-1*S2 - 2*q(mu)*i_*m^2*
[np+nq]^-1*V4 + 2*q(mu)*i_*m^2*[np+nq]^-1*mg^2*S1 + 2*q(mu)*i_*m^2*
[np+nq]^-1*mg^2*V2 )

+ g_(1,p) * (  - 4*p(mu)*i_*T6 - 4*p(mu)*i_*mg^2*V1 + 2*p(mu)*i_*d*T6
- 8*q(mu)*i_*V3 - 4*q(mu)*i_*T7 - 4*q(mu)*i_*mg^2*S1 - 4*q(mu)*i_*
mg^2*V2 + 2*q(mu)*i_*d*V3 + 2*q(mu)*i_*d*T7 )

+ g_(1,q) * ( 4*p(mu)*i_*V3 - 4*p(mu)*i_*T7 + 2*p(mu)*i_*d*T7 - 4*q(mu)
*i_*V4 - 4*q(mu)*i_*T8 + 2*q(mu)*i_*d*V4 + 2*q(mu)*i_*d*T8 )

+ g_(1,mu,q,n) * ( i_*m^2*[np]^-1*S2 - i_*m^2*[np]^-1*mg^2*S1 + i_*m^2*
[np+nq]^-1*S2 - i_*m^2*[np+nq]^-1*mg^2*S1 )

+ g_(1,mu,q,p) * ( 6*i_*V3 + 2*i_*mg^2*S1 - i_*d*V3 )

+ g_(1,mu,q) * (  - 4*i_*M*S2 - 2*i_*M*mg^2*S1 + i_*d*M*S2 )

+ g_(1,mu) * (  - 4*i_*T5 - 2*i_*M^2*S2 - 2*i_*M^2*mg^2*S1 - 2*i_*m^2*
S2 - 2*i_*m^2*mg^2*S1 + 4*i_*d*T5 + 2*i_*d*mg^2*T1 + i_*d*M^2*S2 + i_
*d*m^2*S2 - i_*d^2*T5 + 2*p.p*i_*T6 + 2*p.p*i_*mg^2*T2 - p.p*i_*d*T6
- 4*p.q*i_*V3 + 4*p.q*i_*T7 + 4*p.q*i_*mg^2*T3 - 2*p.q*i_*d*T7 + 2*
q.q*i_*V4 + 2*q.q*i_*T8 + 2*q.q*i_*mg^2*T4 - q.q*i_*d*V4 - q.q*i_*d*
T8 )

+ gi_(1) * ( 4*p(mu)*i_*M*mg^2*V1 - 2*p(mu)*i_*d*M*V3 + 4*q(mu)*i_*M*S2
+ 4*q(mu)*i_*M*mg^2*S1 + 4*q(mu)*i_*M*mg^2*V2 - 2*q(mu)*i_*d*M*S2 - 
2*q(mu)*i_*d*M*V4 );
\end{verbatim}
\section{Appendix C}
Here we list various useful identities.
\begin{eqnarray}
	V_{6} -V_{5} +S_{3} -S_{4} =2 ( T_{6} p.q+T_{7} q.q ) +V_{3} q.q &  & \\
	-V_{5} =2 ( T_{5} +T_{7} p.q+T_{8} q.q ) +q.q V_{4} &  & \\
	I_{3} ( p ) =I_{4} ( p' ) +q^{2} S_{1} +2V_{1} q.p+2V_{2} q^{2}\\
	S_{4} ( p'^{2} ) -S_{3} ( p^{2} ) =-2 ( V_{3} p.q+V_{4} q.q ) -q^{2} S_{2}
\end{eqnarray}

These identities are true for any $p_\mu,q_\mu$.
From them we can derive identities on shell at $q^2=0$.

Consider an example:
\[ I_{3} ( p ) =I_{4} ( p' ) +q^{2} S_{1} +2V_{1} q.p+2V_{2} q^{2} \]
\begin{eqnarray*}
	0= \frac{\partial I_{4} ( p' )}{\partial q_{\mu}} +2q_{\mu} S_{1} +q^{2}
	S_{1, \mu} +2V_{1} p_{\mu} +2V_{1, \mu} q.p+4V_{2} q_{\mu} +2V_{2, \mu}
	q^{2} &  & 
\end{eqnarray*}
Put $q_{\mu} =0$
\begin{eqnarray*}
	0= \frac{\partial I_{4} ( p' )}{\partial q_{\mu}} |_{q_{\mu} =0} 
	+2V_{1} ( 0 ) p_{\mu} 
\end{eqnarray*}
We get the identity:
\begin{equation}
\bar{I}_{3}' + \bar{V}_{1} ( 0 ) =0 
\end{equation}
Remember that $I_{4} ( p'^{2} ) =I_{3} ( p^{2} =p'^{2} )$

On shell we get the following identities:
\begin{eqnarray}
	- \bar{V}_{3} ( q^{2} ) +2 \bar{V}_{4} ( q^{2} ) - \bar{S}_{2} ( q^{2} ) =0
	&  & \\
	\bar{S}_{1} ( q^{2} ) +2 \bar{V}_{2} ( q^{2} ) - \bar{V}_{1} ( q^{2} ) =0 & 
	& 
\end{eqnarray}
\begin{eqnarray}
	\bar{T}_{6} ( 0 ) =- \bar{V}_{5}' - \bar{S}_{4}' &  & \\
	- \bar{V}_{3}(0) +2 \bar{V}_{4}(0) - \bar{S}_{2}(0) =0 &  & \\
	- \bar{V}_{5} =2 \bar{T}_{5} ( 0 )
\end{eqnarray}
\section{Appendix D }
Here we list the small $\lambda$ expansion of the integrals that appear in the calculation of the anomalous magnetic moment.

\begin{eqnarray*}
	\bar{V}_{6} =- \frac{i}{( 4 \pi )^{2}} \left[ \frac{1}{\varepsilon} -
	\frac{\gamma}{2} - \int d x x  \log \left( \frac{M_{e}^{2} x^{2}
		+m_{\gamma}^{2} ( 1-x )}{4 \pi \mu^{2}} \right) \right] = &  & \\
	\begin{array}{l}
		\frac{i}{( 4 \pi )^{2}} \left[ - \frac{1}{\varepsilon} + \frac{\gamma}{2}
		+ \frac{1}{2} \log ( M_{e}^{2} ) + \int d x x  \log \left( \frac{x^{2} +
			\lambda^{2} ( 1-x )}{4 \pi \mu^{2}} \right) \right] \sim\\
		\frac{i}{( 4 \pi )^{2}} \left[ - \frac{1}{\varepsilon} + \frac{\gamma}{2}
		+ \frac{1}{2} \log \left( \frac{M_{e}}{4 \pi \mu^{2}}^{2} \right) -
		\frac{1}{2} \right]
	\end{array} &  & 
\end{eqnarray*}
\begin{eqnarray*}
	\bar{S}_{3} = \frac{i}{( 4 \pi )^{2}} \left[ \frac{2}{\varepsilon} - \gamma
	- \int d x   \log \left( \frac{M_{e}^{2} x^{2} +m_{\gamma}^{2} ( 1-x )}{4
		\pi \mu^{2}} \right) \right] = &  & \\
	\begin{array}{l}
		\bar{S}_{3} =- \frac{i}{( 4 \pi )^{2}} \left[ - \frac{2}{\varepsilon} +
		\gamma + \log \left( \frac{M_{e}^2}{4 \pi \mu^{2}} \right) + \int d x  
		\log ( x^{2} + \lambda^{2} ( 1-x ) ) \right] \sim\\
		- \frac{i}{( 4 \pi )^{2}} \left[ - \frac{2}{\varepsilon} + \gamma + \log
		\left( \frac{M_{e}^2}{4 \pi \mu^{2}} \right) -2 \right]
	\end{array} &  & 
\end{eqnarray*}
\begin{eqnarray*}
	\bar{S}_{3}' = \frac{-i}{( 4 \pi )^{2}} \int d x  \frac{( x^{2} -x
		)}{M_{e}^{2} x^{2} +m_{\gamma}^{2} ( 1-x )} = \frac{-i}{( 4 \pi )^{2}
		M_{e}^{2}} \int d x  \frac{( x^{2} -x )}{x^{2} + \lambda^{2} ( 1-x )} \sim &
	& \\
	\frac{-i}{( 4 \pi )^{2} M_{e}^{2}} ( \log ( \lambda ) +1 ) &  & 
\end{eqnarray*}
\begin{eqnarray*}
	\bar{V}_{6}' = \frac{i}{( 4 \pi )^{2}} \int d x  \frac{x ( x^{2} -x
		)}{M_{e}^{2} x^{2} +m_{\gamma}^{2} ( 1-x )} = \frac{i}{( 4 \pi )^{2}
		M_{e}^{2}} \int d x  \frac{x ( x^{2} -x )}{x^{2} + \lambda^{2} ( 1-x )} \sim
	&  & \\
	\frac{i}{( 4 \pi )^{2} M_{e}^{2}} ( -7.5 ) &  & \\
	\bar{V}_{2} ( 0 ) = \frac{1}{2} \frac{i}{( 4 \pi )^{2}} \frac{1}{M_{e}^{4}}
	\int_{0}^{1} d z \frac{z-1}{z^{2} + \lambda^{2} ( 1-z )} \sim
	\frac{1}{2} \frac{i}{( 4 \pi )^{2}} \frac{1}{M_{e}^{4}} \left( - \frac{\pi}{2
		\lambda} - \log   ( \mathit{\lambda} ) + \frac{1}{2} \right)
\end{eqnarray*}
\begin{eqnarray*}
	&  & \bar{S}_{2} ( 0 ) =- \frac{i}{( 4 \pi )^{2}} \int d x \frac{  ( 1-x
		)}{( M_{e}^{2}   ( x-1 )^{2} +m_{\gamma}^{2} x )} =- \frac{i}{( 4 \pi )^{2}
		M_{e}^{2}} \int d x \frac{  ( 1-x )}{( ( x-1 )^{2} + \lambda^{2} x )} \sim\\
	&  & \frac{i}{( 4 \pi )^{2} M_{e}^{2}} \log ( \lambda )
\end{eqnarray*}
\begin{eqnarray*}
	&  & \bar{T}_{6} ( 0 ) = \frac{-i}{( 4 \pi )^{2} M_{e}^{2}} \int d x 
	\frac{x^{2} ( 1-x )}{( x-1 )^{2} + \lambda^{2} x^{}} \sim\\
	&  & \frac{-i}{( 4 \pi )^{2} M_{e}^{2}} \left( - \log ( \lambda ) -
	\frac{3}{2} \right) = \frac{i}{( 4 \pi )^{2} M_{e}^{2}} \left( \log (
	\lambda ) + \frac{3}{2} \right)
\end{eqnarray*}
\begin{eqnarray*}
	&  & \bar{V}_{3} ( 0 ) =- \frac{i}{( 4 \pi )^{2} M_{e}^{2}} \int d x
	\frac{x ( 1-x )}{( x-1 )^{2} + \lambda^{2} x^{}} \sim\\
	&  & - \frac{i}{( 4 \pi )^{2} M_{e}^{2}} ( - \log ( \lambda ) -1 ) =
	\frac{i}{( 4 \pi )^{2} M_{e}^{2}} ( \log ( \lambda ) +1 )\\
\end{eqnarray*}
\begin{eqnarray}	
	\bar{T}_{5} ( 0 ) =
	- \frac{i}{( 4 \pi )^{2}} \frac{1}{2} \left( - \frac{1}{\varepsilon} +
	\frac{\gamma}{2} + \frac{1}{2} \log \left( \frac{M_{e}^{2}}{4 \pi \mu^{2}}
	\right) + \int d x ( 1-x ) \log ( ( x-1 )^{2} + \lambda^{2} x ) \right)
	\sim \nonumber \\
	- \frac{i}{( 4 \pi )^{2}} \frac{1}{2} \left( - \frac{1}{\varepsilon} +
	\frac{\gamma}{2} + \frac{1}{2} \log \left( \frac{M_{e}^{2}}{4 \pi \mu^{2}}
	\right) - \frac{1}{2} \right)
\end{eqnarray}
\section{Appendix E}

To avoid the ambiguity in the prescription we notice that the photon self
energy satisfies the Ward identity automatically, after evaluating the $\bar{n}_\mu->0$ limit. It is because we have traces
of gamma matrices, thus pure numbers. 

So, we introduce a basis of gamma functions and expand any matrix function $A$ in this basis:
\begin{eqnarray*}
	1, \gamma_{\mu}, \sigma_{\mu \nu}, \ldots . &  & \\
	A = trace(A) 1 + trace(A\gamma_\alpha)
	\gamma_{\alpha} + \ldots . &  & 
\end{eqnarray*}
All traces are functions of $n_\mu,\bar{n}_\mu$ and external momenta. To obtain the limit  $\bar{n}_\mu \rightarrow 0$, we must proceed in the following way. First apply in all monomials the identities $n.n=\not{n}.\not{n}=0$, $n.\not{n}=1$. Afterwards put $\bar{n}_\mu=0$ everywhere.

Moreover the identity exploited in equation (7.65) of \cite{Peskin} will be applied to each trace. Due to the linear and cyclic properties of the trace, the proof of the Ward-Takahashi  identity will go through. Therefore the trace method always lead to a gauge invariant, $Sim(2)$ invariant result.

We have checked that in VSR QED the trace method gives the same results as the method explained in chapter 4.
\section{Appendix F:Feynman rules}
To draw the Feynman graphs we used \cite{ellis}
\begin{figure}[h]
	\includegraphics[scale=0.4]{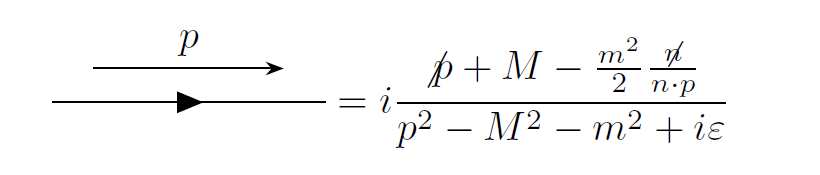}
	\includegraphics[scale=0.4]{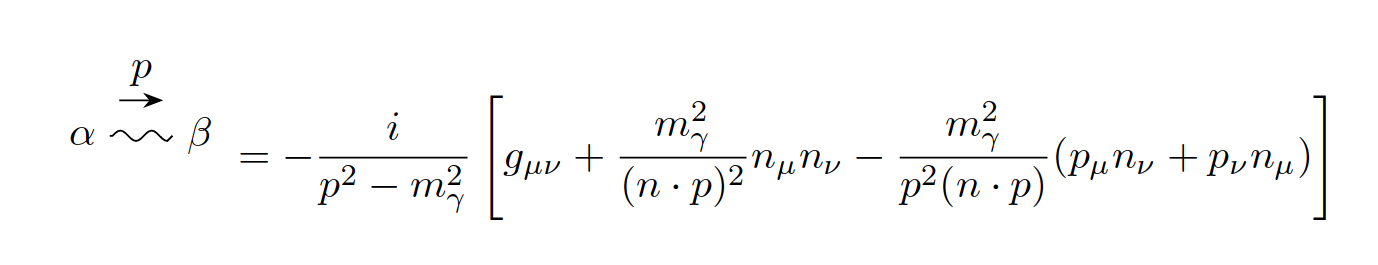}
	\includegraphics[scale=0.4]{v3.png}
	\includegraphics[scale=0.4]{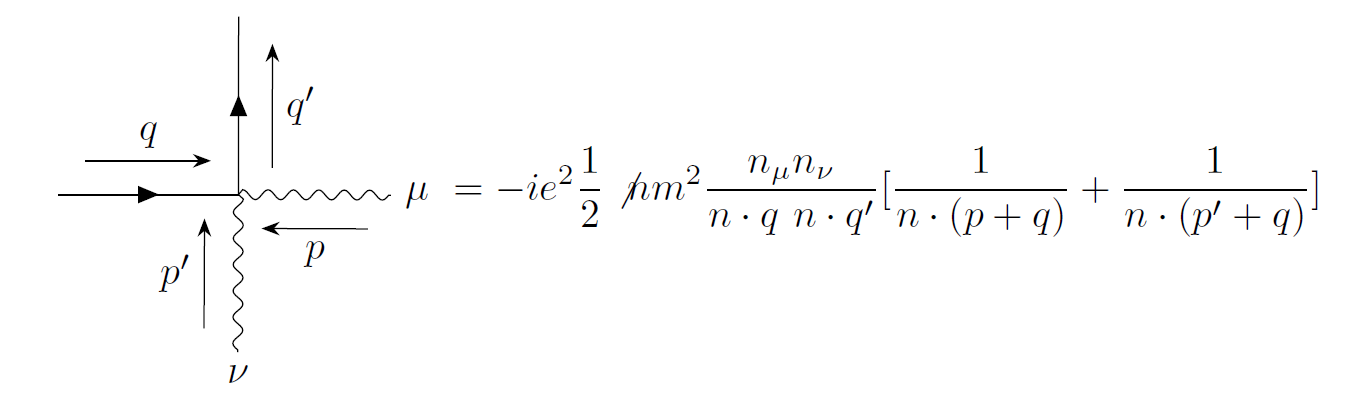}
	\caption{Feynman rules for one loop computations:electron propagator,photon propagator,$A_\mu ee$ and $A_\mu A_\nu ee$ vertex.}
	\label{Fig: Fey rules}
\end{figure}


\begin{thebibliography}{99}
	\bibitem{w2}The CMS collaboration, "Evidence for the direct decay of the
125 GeV Higgs boson to fermions", Nature Physics 10, 557$-$560 (2014).

\bibitem{Langacker}Paul Langacker. The Standard model and Beyond. CRC
Press, A Taylor and Francis Group (2010).

\bibitem{mohapatra}Rabindra Mohapatra. Unification and Supersymmetry: The
Frontiers of Quark-Lepton Physics, Third Edition. Springer (2002).

\bibitem{CG1}A. G. Cohen and S. L. Glashow, Very special relativity,
Phys.Rev.Lett. 97 (2006) 021601.

\bibitem{CG2}Cohen, A. and Glashow, S., "A Lorentz-Violating Origin of
Neutrino Mass?", hep-ph 0605036.

\bibitem{ja1}Alfaro,J,Gonz{\'a}lez,P and {\'A}vila,R,Phys Rev. D91(2015)
105007,Addendum:Phys. Rev. D91(2015) no. 12,129904.

\bibitem{plb} J. Alfaro, A Sim(2) invariant dimensional regularization, Phys. Lett. B 772, 100 (2017)
\bibitem{AUniverse} J. Alfaro, Feynman rules, ward identities and loop corrections in very special relativity Standard Model, Universe 5, 16 (2019).

\bibitem{AS1} J. Alfaro and A. Soto, Phys. Rev. D 100, 055029 DOI:https://doi.org/10.1103/PhysRevD.100.055029	
\bibitem{AS2} Jorge Alfaro and Alex Soto. Schwinger model a la Very Special Relativity. Phys. Lett.B, 797:134923, 2019.
\bibitem{Mandelstam}S. Mandelstam, Nucl. Phys. B213, 149 (1983).
\bibitem{Leibbrandt}G. Leibbrandt, Phys. Rev. D29, 1699 (1984).
\bibitem{AML}J. Alfaro, Mandelstam-Leibbrandt prescription, Phys. Rev. D 93, 065033 (2016); Erratum, 94, 049901(E) (2016).
\bibitem{ASan}J.Alfaro and A.Santoni,PLB 829(2022)137080;
 https://doi.org/10.1016/j.physletb.2022.137080
\bibitem{ASanS} A. Santoni,J. Alfaro and A. Soto, in preparation.
\bibitem{form} FORM:J.A.M.Vermaseren,New features of FORM.  math-ph/0010025.
\bibitem{Pokorski}  S. Pokorski, Gauge Field Theories, Cambridge
Monographs on Mathematical Physics (Cambridge
University Press, Cambridge, England, 2000).
\bibitem{Peskin}An Introduction to Quantum Field Theory. M. Peskin, and D. Schroeder. Westview Press, (1995 )Reading, USA: Addison-Wesley (1995)
\bibitem{pdg} R.L. Workman et al. (Particle Data Group), Prog. Theor. Exp. Phys. 2022, 083C01 (2022)
\bibitem{ammexp}Hanneke, D.; Fogwell Hoogerheide, S.; Gabrielse, G., Physical Review A. 83 (5): 052122. doi:10.1103/PhysRevA.83.052122. S2CID 16902741
\bibitem{ammtheo} Aoyama, Tatsumi; Hayakawa, Masashi; Kinoshita, Toichiro; Nio, Makiko ,Physical Review D. 91 (3): 033006.doi:10.1103/PhysRevD.91.033006
\bibitem{katrin} The KATRIN Collaboration. Direct neutrino-mass measurement with sub-electronvolt sensitivity. Nat. Phys. 18, 160?166 (2022). https://doi.org/10.1038/s41567-021-01463-1
\bibitem{cosmo1} Alam, S. et al. Completed SDSS-IV extended baryon oscillation spectroscopic
survey: cosmological implications from two decades of spectroscopic surveys
at the Apache Point Observatory. Phys. Rev. D 103, 083533 (2021).
\bibitem{cosmo2} Aghanim, N. et al. Planck 2018 results. VI. Cosmological parameters. Astron.Astrophys. 641, A6 (2020).
\bibitem{fermilab}B. Abi et al. Muon $g-2$, Phys. Rev. Lett. 126 (2021) 141801 [arXiv:2104.03281 [hep-ex]].
\bibitem{brook}G. W. Bennett et al. Muon $g-2$, Phys. Rev. D 73 (2006) 072003 [arXiv:hep-ex/0602035
[hep-ex]].
\bibitem{SMmuon}T. Aoyama et al., Phys. Rept. 887 (2020) 1 [arXiv:2006.04822 [hep-ph]].
\bibitem{ellis} Joshua Ellis. 'TikZ-Feynman: Feynman diagrams with TikZ'. arXiv: 1601.05437 [hep-ph]

\end{thebibliography}
\end{document}